\documentclass{ar-1col-S2O}

\usepackage[comma]{natbib}
\usepackage{url}
\usepackage{lscape}

\newcommand\ms{$\mathrm{m\ s^{-1}}$}
\newcommand\kms{$\mathrm{km\ s^{-1}}$}

\newcommand{\Mpl}{\ensuremath{M_{p}}}

\newcommand{\Mstar}{\ensuremath{M_{\star}}}

\jname{To appear in Annual Reviews of Astronomy \& Astrophysics}
\jvol{}
\jyear{2025}

\begin{document}

\markboth{Burt, Dumusque, \& Halverson}{Precise Radial Velocities}

\title{Precise Radial Velocities}

\author{Jennifer A. Burt$^1$, Xavier Dumusque$^2$, and Samuel Halverson$^1$
\affil{$^1$Jet Propulsion Laboratory, California Institute of Technology, 4800 Oak Grove Drive, Pasadena, CA 91109, USA}
\affil{$^2$Astronomy Department of the University of Geneva, 51 ch. de Pegasi, 1290, Versoix, Switzerland}}

\begin{abstract}
Precise measurements of a star's radial velocity (RV) made using extremely stable, high resolution, optical or near infrared spectrographs can be used to determine the masses and orbital parameters of gravitationally-bound extra-solar planets (exoplanets). Indeed, RV surveys and follow up efforts have provided the vast majority of published exoplanet mass measurements and in doing so have enabled studies into exoplanet interior and atmospheric compositions. Here we review the current state of the RV field, with particular attention paid to:
\begin{itemize}
\item The evolution of precise RV methodologies over the past two decades
\item Modern RV spectrograph designs that can be calibrated to a stability \\ level of $\leq$ 50 c\ms\ over timescales of years
\item RV data reduction and post-processing techniques that minimize the \\ impact of instrument systematics and stellar variability 
\item Techniques for detecting exoplanets in RV data and disentangling \\ planetary signals from stellar variability
\end{itemize}
\end{abstract}

\begin{keywords}
spectroscopic techniques, exoplanets, radial velocities, high resolution spectrographs, stellar variability
\end{keywords}
\maketitle

\tableofcontents

\section{INTRODUCTION}
The Doppler radial velocity (RV) method of detecting exoplanets has played a foundational role in our efforts to discover, characterize, and understand exoplanets in the solar neighborhood. RV surveys dominated the early days of exoplanet science, producing 99 of the first 100 exoplanet discoveries around sun like stars and revealing previously unknown classes of planets such as hot Jupiters \citep{MayorQueloz1995}. In total, RV detections account for almost 20\% of the 6000+ exoplanets that have been confirmed to date \citep{ExoplanetArchive2025} and looking forward, Extreme Precision RV (EPRV) currently offers the best chance to detect Earth analogs; rocky planets orbiting within the habitable zones of sun like stars.  

In this review we aim to provide the reader with an synopsis of the modern precise radial velocity field, focusing on the underlying techniques that determine our Doppler exoplanet detection capabilities rather than on the exoplanet discoveries themselves. The sections that follow do not provide an exhaustive summary of the RV field, but instead highlight key advancements in our instruments, data reduction pipelines, and exoplanet/stellar variability modeling tools that have driven the field to its current state. 

\section{OVERVIEW OF THE RADIAL VELOCITY LANDSCAPE THROUGH TIME} 

\subsection{Keplerian Orbits \& Doppler Spectroscopy}\label{sec:Orbits}

Gravitational interactions between a host star and its planet cause both objects to trace out elliptical orbits with the barycenter of the system at one focus. These ellipses can be described using seven parameters: the semi-major axis ($a$), the eccentricity ($e$), the orbital period ($P$), the time of periastron ($t_{p}$), the orbital inclination ($i$), the longitude of the ascending node ($\Omega$), and the argument of periastron ($\omega$). In a two-body system the star and the planet share the same orbital period and eccentricity but have opposite arguments of periapsis, meaning that they sit on opposite sides of their system barycenter (Figure 1). 

\begin{marginnote}[]
\entry{barycenter}{center of mass of two or more bodies that orbit one another, also the point about which the bodies orbit.}
\end{marginnote} 

If the star's orbital ellipse is oriented such that it is even partially edge-on as seen by an observer then its repeated motion along that path will take it cyclically towards and away from the observer and produce corresponding cyclical shifts in the radial component of the star's velocity. The radial component of that elliptical motion over time is 

\begin{equation}
V_{r}(t) = K[\cos(\nu(t)+\omega) +e\cos(\omega) ]
\end{equation}

where $\nu(t)$ is the true anomaly and $K$ is the RV semi-amplitude, given by

\begin{equation}
K = \frac{2\pi\ a\ \sin(i)}{P (1-e^{2})^{1/2}}
\end{equation}

Following \citealt{Cumming1999}, $K$ can be rewritten to reference the masses of the star and planet (\Mstar\ and \Mpl, respectively) rather than the semi-major axis $a$ by substituting in a general form of Kepler's third law:

\begin{equation}\label{eqn:semiamp}
K = \left(\frac{2\pi G}{P}\right)^{1/3} \frac{\Mpl \sin(i)}{(\Mstar + \Mpl)^{2/3}} \frac{1}{(1-e^{2})^{1/2}}
\end{equation}

\begin{figure}[h]
\includegraphics[width=4.5in]{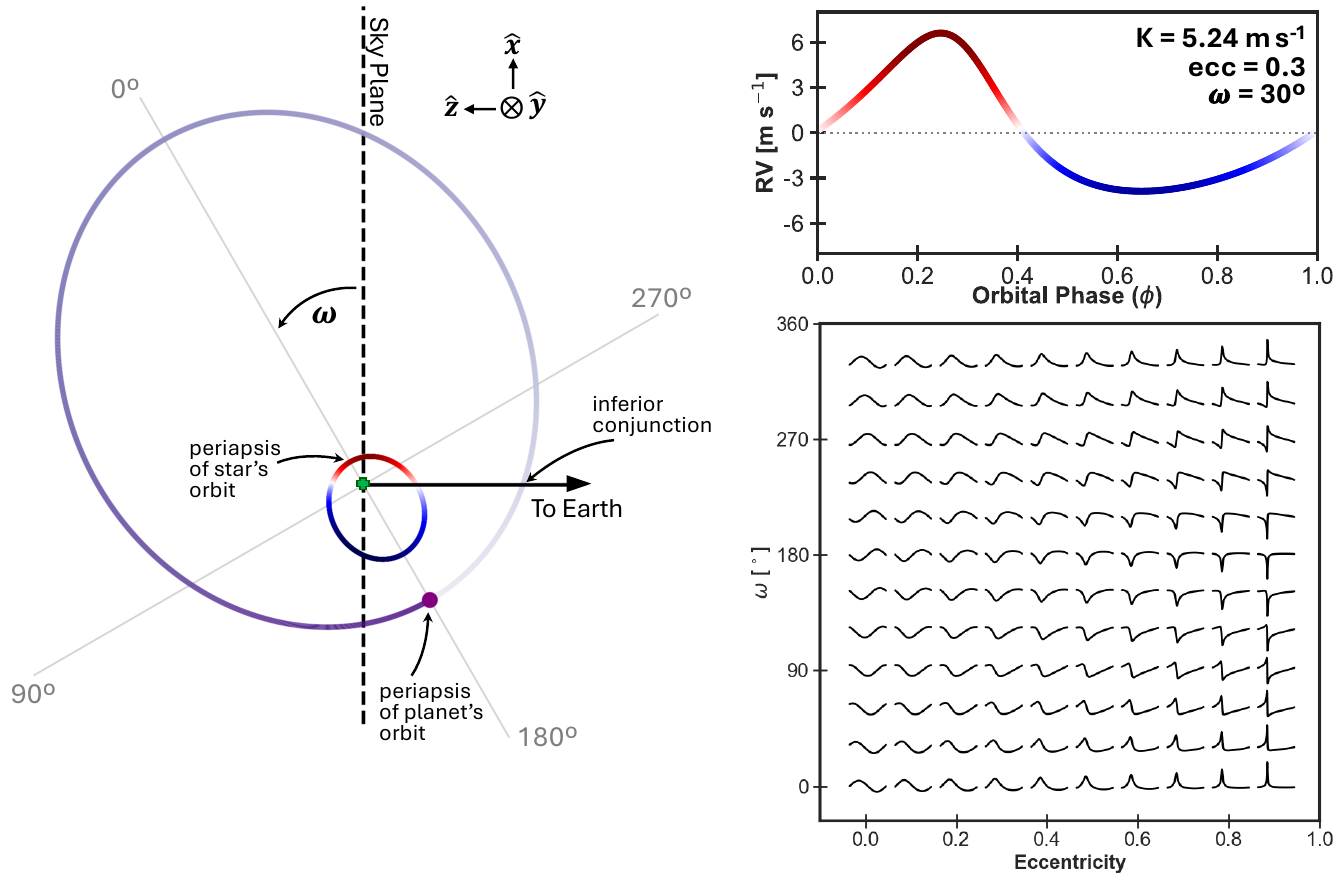}
\caption{Left: Schematic representation of the two-body Keplerian orbit of an exoplanet (in purple) and its host star (color coded with star's red/blue shift as observed from Earth), scale exaggerated for clarity. The two orbits have the same period and eccentricity, but their semi-major axes are scaled by the star-planet mass ratio. The star and planet are always on opposite sides of their barycenter (green cross) such that when the planet is moving away from an observer, the star is moving towards the observer and vice versa. Top Right: RV curve corresponding to the stellar orbit depicted on the left. Bottom Right: RV curves can exhibit a wide range of shapes based on their orbital eccentricity and longitude of periastron ($\omega$) values. Higher orbital eccentricity produces a more peaked curve and a higher maximum semi-amplitude, with most of the stellar velocity shift taking place as the star passes through periapsis. Changes in the orbit's orientation relative to the observer ($\omega$) shifts the phase in the orbit where the most significant RV variations occur.}
\label{fig:RV_Orbits}
\end{figure}

Thus if we are able to measure the star's radial velocity over time and identify the Keplerian model that best describes the resulting RV measurements, then the best-fit orbital parameters can be combined with knowledge of the mass of the host star to derive the mass of the planet. Depending on the eccentricity of the orbit and its alignment to the observer, the Keplerian model can exhibit a wide range of shapes (Figure \ref{fig:RV_Orbits}). 

An important note is that the planet mass term, $M_{p}$, is degenerate with the planet's orbital inclination, a parameter than can only be determined via transit photometry \citep{Winn2010} or in conjunction with astrometric observations \citep[see, e.g.,][]{Kiefer2019}. What we determine via RV-only detections is actually the \emph{minimum mass} of the planet or $M_{p}\sin(i)$. The further the planet is from edge-on (defined as $i = 90^\circ$) the larger its true mass will be, but this is a relatively slow function of inclination and given reasonable assumptions about stellar inclinations in the galaxy we expect $\sim$87\% of RV-detected planets to have a true mass that is at most two times their $M_{p}\sin(i)$ value.

The radial velocity of a star can be measured via centroid shifts in the star's spectral absorption lines, as described by the Doppler Equation

\begin{equation}
z = \frac{\lambda - \lambda_{o}}{\lambda_{o}} = \frac{1 + V_{r}/c}{\sqrt{1 - V^{2}/c^{2}}}-1
\end{equation}

where z is the red-shift, $\lambda$ is the line's observed wavelength, $\lambda_{o}$ is its rest wavelength, $V_{r}$ is the corresponding radial velocity, c is the speed of light, and $V$ is the total velocity relative to the observer. Positive z and $V_{r}$ values generally correspond to the object moving away from the observer, which we refer to as a red-shift, while negative z and $V_{r}$ values represent motion towards the observer and are referred to as blue-shifts. 

\subsection{The Evolution of Precise RV Measurements}

Typical exoplanets can induce RV changes ranging from $\pm$100 of \ms\ for hot Jupiters down to $\pm$10 c\ms\ for Earth analogs on their stars. RV instruments operate in the visible (400 - 900nm) or near infrared (1-2$\mu$m) regimes and utilize detectors with pixel scales of roughly 1 km s$^{-1}$ pixel$^{-1}$, so planet-induced Doppler shifts have typical scales of only 0.1 to 0.0001 pixels. Such shifts cannot be precisely measured using a single absorption line, and must instead be determined via a \lq{}bulk\rq{} RV measurement that combines the signal of hundreds or thousands of absorption lines within a stellar spectrum. 

\begin{figure}[h]
\includegraphics[width=4.5in]{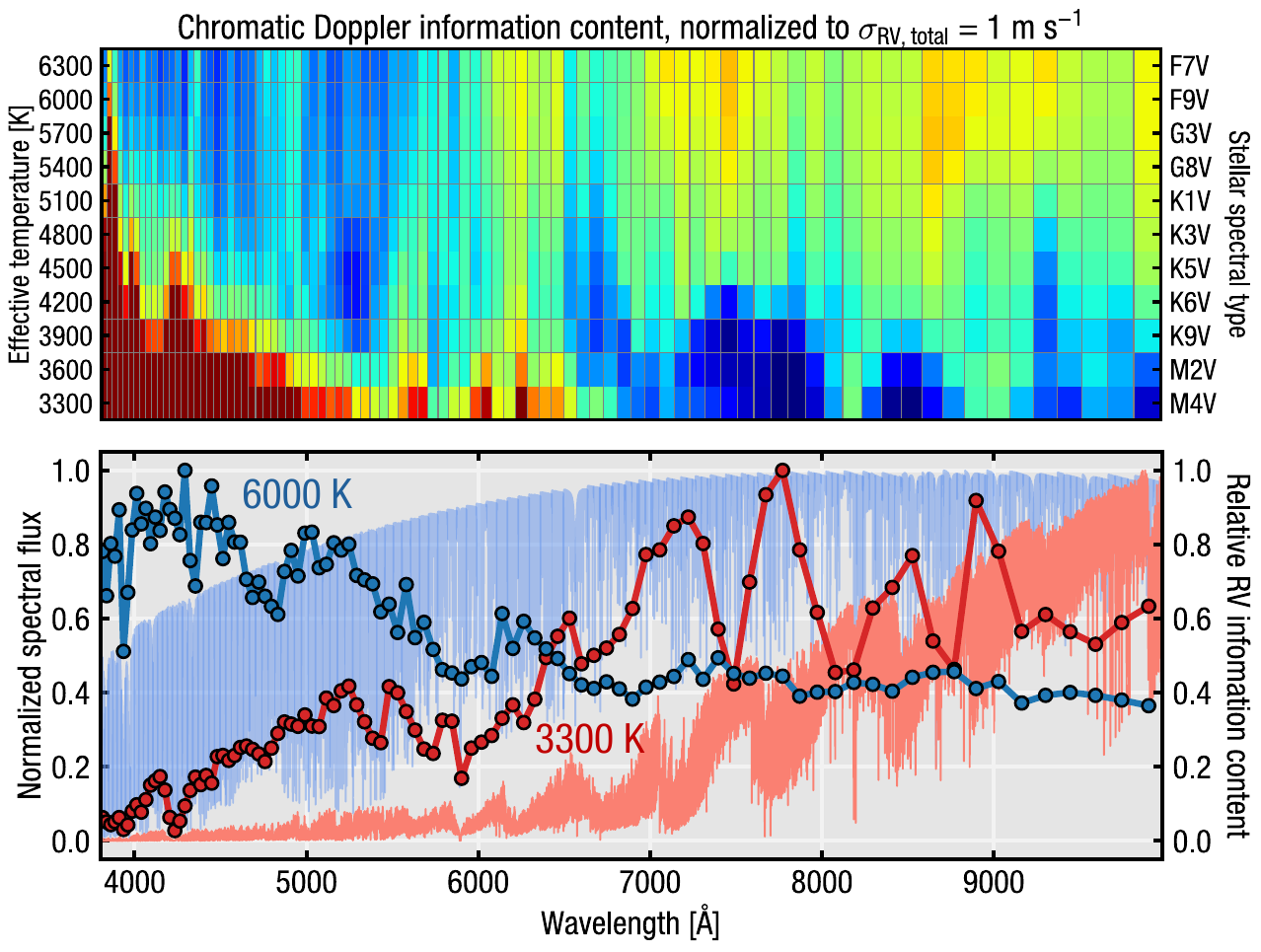}
\caption{Top: Relative Doppler radial velocity (RV) information content as a function of spectral type and wavelength region normalized to $\sigma_{\mathrm{RV, total}}$ = 1 \ms, following the quality-factor (Q) calculations detailed in \citealt{Bouchy2001}. Bluer colors indicate higher RV information content levels and correspondingly smaller contributions to the total RV uncertainty. Monochromatic color bins denote individual grating diffraction orders. Bottom: Stellar flux (transparent spectra, left-axis) and relative Doppler information content ($1/\sigma_\mathrm{RV}$, solid points and lines, right axis) as a function of wavelength for simulated Sun-like (6000 K) and M-dwarf (3300 K) spectra recorded from a notional ground-based RV system. These curves effectively represent \lq{}slices\rq{} of the 2D map above.}
\label{fig:RVInformationContent}
\end{figure}

Quantitatively, the Doppler information contained within a spectrum is highly dependent on the characteristics of the spectral lines. For example, deeper and narrower absorption lines yield more RV information because their central positions, which are used to measure the Doppler shift, can be more precisely determined. Efforts to formalize the RV information content of a stellar spectrum and establish the fundamental photon noise limits of RV measurements started with the definition of an intrinsic spectral quality factor \citep[$Q$,][]{Connes1985}. This was later expanded to consider the impact of the spectral range and resolution of a given instrument and the rotational broadening of the stellar spectrum \citep{Bouchy2001} as well as the host star's surface gravity, metallicity, and macroturbulence \citep{BeattyGaudi2015}. Figure \ref{fig:RVInformationContent} depicts relative RV information content as a function of wavelength and spectral type.

\begin{figure}[ht]
\includegraphics[width=4.5in]{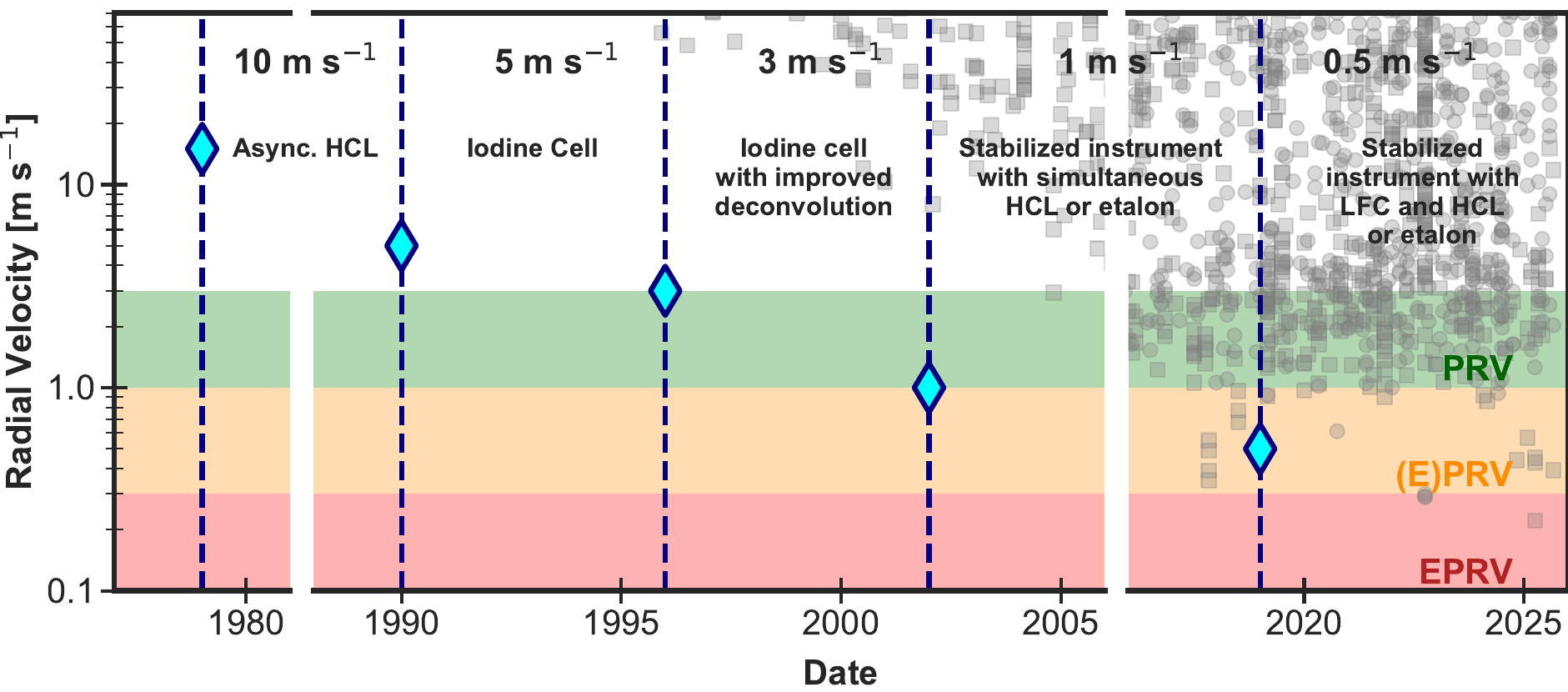}
\caption{Major advancements in radial velocity (RV) precision over time (cyan diamonds) and the semi-amplitudes of planets discovered each year. RV-only detections are depicted as squares, while planets detected first in transit and then followed up with RV facilities are depicted as circles. The stellar activity \lq{}barrier\rq{} has made the discovery of planets with RV semi-amplitudes below 1 \ms\ extremely challenging, but in recent years the combination of modern RV spectrographs, improved post-processing pipelines, and more nuanced stellar activity mitigation techniques has produced a growing number of planets in this K $\leq$ 1 \ms\ regime.}
\label{fig:precision_time}
\end{figure}

Given the need to resolve and measure the shifts of thousands of lines across hundreds of nanometers of wavelength range, RV instruments are built around very high resolution  spectrographs that cross-disperse incoming light into dozens of diffraction orders. Key to this endeavor is the precise determination of a corresponding wavelength solution for each spectrum, in order to measure the absorption lines' central wavelengths to high fidelity. In the late 1980s and early 1990s hollow cathode lamps (HCLs) containing well characterized gases became standard wavelength calibrators, but these were generally observed significantly before or after the target stars. This temporal offset led to flexures within the telescope and/or spectrograph between the observations and discrepant illumination of the optics which limited the RV precision to $\sim$200 \ms\ \citep{Latham1989, Marcy1989, Duquennoy1991}. Advancement to the era of `precise' radial velocities, which here we define as individual RV precision at the $\leq$ 1-2 \ms\ level, started in the late 1990s via improvements to the wavelength calibration techniques for iodine cell instruments \citep{Butler1996} and then with the development of stabilized spectrographs with simultaneous wavelength calibration sources (Figure \ref{fig:precision_time}).

\begin{figure}[h]
\includegraphics[width=5in]{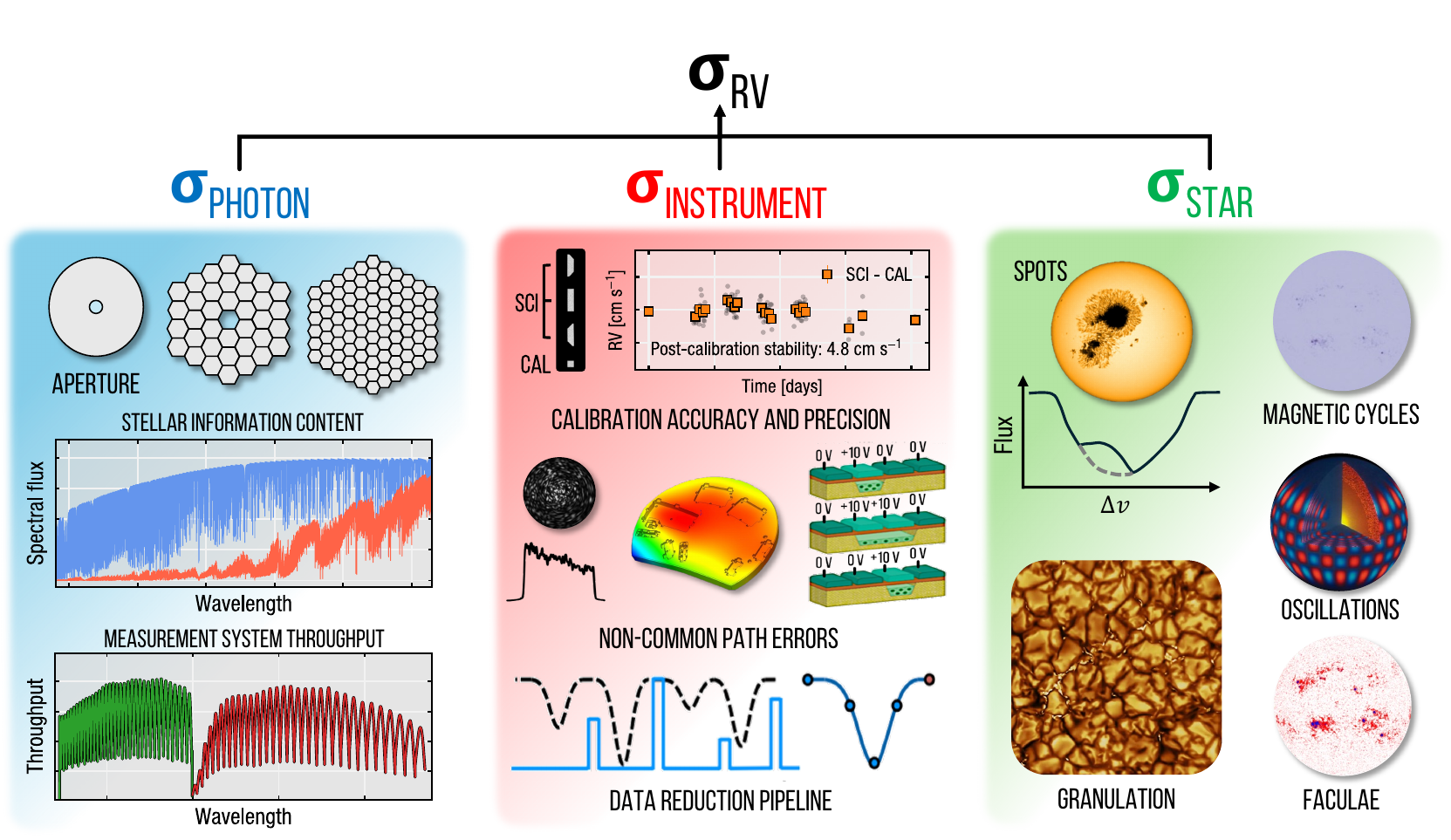}
\caption{Overview of different sources of random and systematic errors that impact Doppler radial velocity (RV) measurements. Left: The number and distribution of photons collected in the stellar spectra dictate the photon-limited Doppler measurement floor \citep{Bouchy2001}. This depends on the parameters of the star being observed (brightness, spectral type, rotational velocity, etc.) and properties of the facility (aperture, atmospheric extinction, instrument efficiency). Middle: Errors associated with instabilities in the measurement system, including both the instrument and the methods used to extract velocity measurements from raw images, directly affect RV performance. Right: Stellar astrophysical noise sources span a range of amplitudes and timescales. These affects manifest at both the spectral and integrated RV level and span timescales from seconds to decades. Sub-figures in each category are inspired by figures and references in \citet{Crass2021}.}
\label{fig:rv_error}
\end{figure}

The RV error for a given measurement is set by three primary components (Figure \ref{fig:rv_error}). First is the photon noise, set by the RV information content of the star, the bandpass of the RV spectrograph used to observe it, and the aperture/throughout of the telescope that the spectrograph is mounted to. Second is the contribution from instrument and facility systematics which can capture residual temperature or pressure fluctuations, instabilities in the wavelength calibration, and errors in the spectrum and/or RV extraction pipelines. And third are astrophysical noise sources from the star itself including line shape deformations from phenomena such as spots, faculae, and granulation. 

\section{THE RADIAL VELOCITY MEASUREMENT SYSTEM} \label{sec:RV_sys} 
The nature of the spectroscopic measurements used for precision Doppler measurements necessitates both tailored hardware and creative analysis techniques. Here we describe the instrumental aspects of modern Doppler spectroscopy. Table \ref{tab:inst_list} provides an overview of the precision RV spectrographs that are currently, or soon to be, in operation along with links to their respective instrument papers. The majority are stabilized instruments which provide more Doppler information content from a given spectrum and simplify the derivation of the RV value at the expense of much more exacting control on the stability of the instrument.

\begin{table}[h]
\tabcolsep7.5pt
\caption{Current and upcoming RV instruments capable of delivering single measurement precision $\leq$1 \ms\ in the visible or $\leq$3 \ms\ in the NIR.}
\begin{center}
\begin{tabular}{@{}l|c|c|c|c|c@{}}
\hline
Instrument & First Light& Bandpass & Resolution & Telescope & Reference \\
 & & [nm] & [$\lambda$ / $\Delta\lambda$] & \& Aperture [m] & \\
\hline
HARPS*	&	2003	&	380-690	&	115,000	& La Silla 3.6m	&	\citealt{Mayor2003}	\\
HARPS-N*	&	2012	&	380-690	&	115,000	& TNG [3.6m]	&	\citealt{Cosentino2014}	\\
SOPHIE+	&	2012	&	380-690	&	75,000	&	OHP 1.93m	&	\citealt{Bouchy2013}	\\
APF	&	2014	&	500-620	&	110,000	&	APF [2.4m]	&	\citealt{Vogt2014}	\\
CARMENES	&	2016	&	520-1710	&	82,000	&	Calar Alto 3.5m	&	\citealt{Quirrenbach2014}	\\
iSHELL	&	2016	&	2180-2470	&	80,000	&	IRTF [3.2m]	&	\citealt{Cale2019}	\\ 
IRD	&	2017	&	970-1750 	&	70,000	&	Subaru [8.2m]	&	\citealt{Kotani2018}	\\
EXPRES*	&	2018	&	390-780	&	137,500	&	LDT [4.3m]	&	\citealt{Blackman2020}	\\
HPF	&	2018	&	800-1270	&	50,000	&	HET [10m]	&	\citealt{Mahadevan2014}	\\
PFS 	&	2018	&	500-620	&	120,000	&	Magellan Clay [6m]	&	\citealt{Crane2008}	\\
ESPRESSO*	&	2018	&	380-790	&	140,000	&	VLT [8m]	&	\citealt{Pepe:2021aa}	\\
PARVI*	&	2019	&	1145-1766	&	60,000	&	Hale [5m]	&	\citealt{Cale2023}	\\
SPIRou	&	2019	&	980-2350	&	64,000	&	CFHT [3.6m]	&	\citealt{Donati2020}	\\
MAROON-X*	&	2020	&	500-920	&	85,000	&	Gemini-N [8m]	&	\citealt{Seifahrt2020}	\\
NEID*	&	2021	&	380-930	&	115,000	&	WIYN [3.4m]	&	\citealt{Schwab2016}	\\
KPF*	&	2022	&	445-870	&	97,000	&	Keck I [10m]	&	\citealt{Gibson2024}	\\
PARAS-2	&	2022	&	380-690	&	107,000	&	PRL [2.5m]	&	\citealt{Chakraborty2024}	\\
NIRPS*	&	2023	&	970-1800	&	84,000	&	La Silla 3.6m	&	\citealt{Artigau2024}	\\
\hline													
HARPS3*	&	2026	&	380-690	&	115,000	&	INT [2.5]	&	\citealt{Thompson2016}	\\
iLocater	&	2026	&	970-1310	&	190,000	&	LBT [8.4m]	&	\citealt{Crass2022}	\\
MARVEL	&	2026	&	380-950	&	135,000	&	Mercator Obs. [0.8m]	&	\citealt{Pember2022}	\\
HISPEC	&	2026	&	980-2500	&	100,000	&	Keck II [10m]	&	\citealt{Konopacky2023}	\\
2ES	&	2027	&	370-890	&	120,000	&	ESO/MPG 2.2m	&	\citealt{Sturmer2024}	\\
G-CLEF	&	2029	&	350-950	&	105,000	&	GMT [25m]	&	\citealt{Szentgyorgyi2018}	\\
\hline
\end{tabular}
\end{center}
\begin{tabnote}
$^{\rm *}$Spectrographs that have, or are soon adding, a solar feed to allow day time observations of the Sun.\\
\end{tabnote}
\label{tab:inst_list}
\end{table}

\begin{figure}[h]
\includegraphics[width=5in]{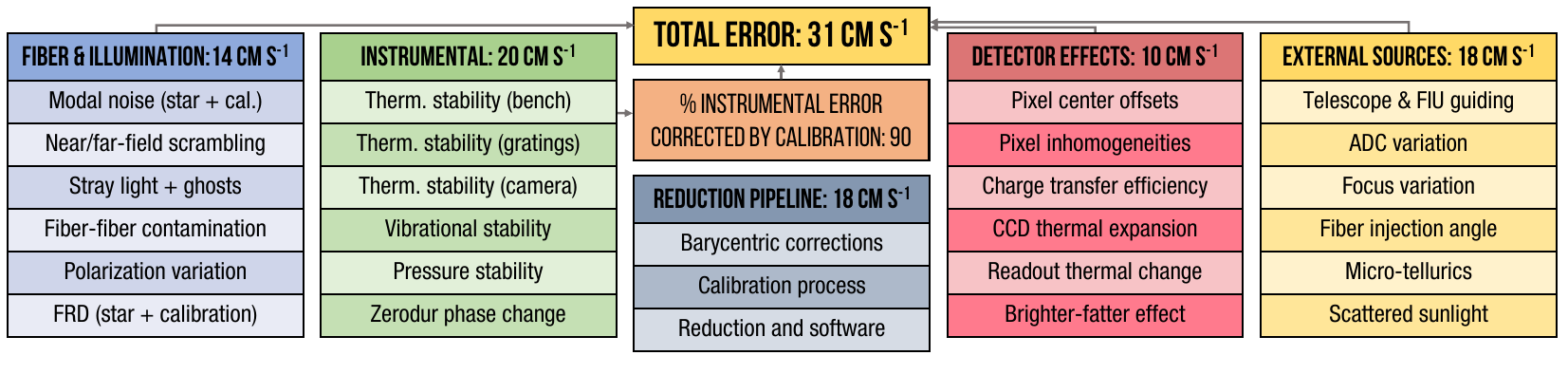}
\caption{Typical components in a precision Doppler radial velocity (RV) single measurement error (measurement system only). Figure adapted from the Keck Planet Finder performance budget \citep{Gibson2024}. Individual error terms are categorized by type and/or instrumental subsystem. Some errors are tracked by simultaneous calibration (listed as \lq{}instrumental\rq{} in the table) and others are not. This follows similar system-level analyses from other EPRV facilities \citep{Podgorski2014, Blackman2020, Bechter2020b, Halverson2016}. Below 1 {\ms}, a multitude of random and systematic errors begin to make measurable contributions to the RV measurement noise floor. These errors have origins ranging from the Earth's atmosphere (tellurics), to the telescope, to the spectrometer, to the methods used to extract the data and compute stellar RVs. While many of these errors are constrainable via analysis techniques or lab tests, others are active areas of research for pushing to the c{\ms} level \citep[e.g.,][]{Blake2017}.}
\label{fig:error_budget}
\end{figure}

Revisiting the scale of the RV measurements can be useful for contextualizing the instrumentation required. Consider a generic spectrometer with a spectral resolving power $R = \lambda/\Delta\lambda = $ 100,000 (typical for current RV instruments, see Table \ref{tab:inst_list}). This translates to an effective 1-dimensional (1D) spectral line spread function (LSF) width of $c$/100,000 = 3 {\kms} in velocity. Assuming a $\sim$3-pixel sampling of the LSF (a reasonable value for most modern RV instruments), this equates to an effective velocity \lq{}dispersion\rq{} of $\sim$1 {\kms} per detector pixel. For an Earth-analog planet, with a 9 c{\ms} RV semi-amplitude, this implies that the instrumental systematics must be controlled and/or calibrated to better than 1/10000th of a pixel over multi-year timescales. Ideally any measurement systematics should be controlled to at least 10$\times$ better than this ($\sim$c{\ms}) to improve statistical significance of detections, shorten survey times, and reduce the complexity of the data extraction process. Developing instruments with intrinsic stability and/or calibratability at this level inherently requires a multitude of separate subsystems to work in concert (see Figure~\ref{fig:error_budget}). Below we review the major instrument design features implemented for precision RV science observations over the past two decades, from light collection to wavelength calibration.

\subsection{Light Injection \& Illumination}
Any spectrometer is fundamentally producing a polychromatic image of the entrance slit. Any variations in the entrance slit illumination directly lead to changes in the LSF shape. More subtly, changes in the spectrometer pupil illumination affect the relative weighting of the optical aberrations that set the instrument point-spread function (PSF). It is therefore critical to stabilize both the slit (\lq{}near-field\rq{}) and pupil (the \lq{}far-field\rq{}) illumination to not introduce systematic errors. RV facilities achieve this with multiple techniques.

First, the image of the star is placed precisely into the fiber entrance aperture, or entrance slit for slit-fed instruments. This typically involves some form of fast tip/tilt system to maintain guiding performance at a level significantly smaller than the instrument entrance slit or fiber. Different instruments employ different approaches to achieve this, ranging from guiding using the reflected light off an entrance slit (HIRES, PFS, APF), to imaging a pin-hole mirror at the telescope focal plane (HARPS, ESPRESSO, MAROON-X) to using an in-band pickoff imaging system (NEID), to guiding out of band completely (KPF). Atmospheric dispersion correctors are used to maximize chromatic coupling efficiency and ensure the stellar image centroid is aligned across all wavelengths. Second, optical fibers are used to spatially homogenize and stabilize the spectrometer illumination. Fibers naturally provide some degree of spatial \lq{}scrambling\rq{}, smoothing over input inhomogeneities in the spatial flux distribution. This is critical for most modern EPRV systems, which rely on non-common-path calibration approaches where any illumination variability in the stellar spectrum is not traced by the wavelength reference channel. While fibers suffer from focal ratio degradation \citep{Ramsey1988} which leads to some throughput loss, they still provide a highly efficient method for delivering light to the spectrometer from the telescope focus.

Injection variations at the fiber entrance produce subtle changes in the near and far-field output. While fibers provide some stabilization of the image, they are imperfect scramblers at the levels required to support $<$1 {\ms} measurements. To combat this, non-circular-core geometries that have demonstrated significantly better near-field stability are used. Fiber double-scramblers are also often employed to homogenize and stabilize both the near and far field patterns. Current state of the art systems pair non-circular fibers and double scramblers to reach illumination stability in both the near and far-field at the 10$^{-4}$ level \citep{Halverson2015, Avila1998, Avila2008, Barnes2010, Bouchy2013, Heacox1986, Sturmer2014, Spronck2012}.

\begin{marginnote}[]
\entry{double-scrambler}{optical device that images the far-field illumination pattern of one fiber onto the face (near-field) of another, and vice versa.} 
\end{marginnote}

A drawback of the multi-mode optical fibers used in most seeing-limited RV systems is the requirement that injected light must be coupled into a finite number of propagation modes \citep{Lemke2011}. These modes are populated with different intensities and phases based on the injection conditions, which naturally vary over time (e.g. due to seeing variations). The interference of these modes leads to a highly structured speckle pattern at the fiber exit, which gives rise to \lq{}modal noise\rq{}. This both imposes imposes a signal-to-noise (S/N) ceiling and introduces apparent centroid variability \citep{Oliva2019, Baudrand2001, Goodman1980, Rawson1980}. The number of populated modes in a fiber goes as $(d/\lambda)^{2}$ where $d$ is the fiber diameter \citep{Lemke2011}. This means modal noise is significantly worse at NIR wavelengths \citep{Oliva2019} and for instruments with small fibers. This issue is particularly challenging for coherent emission spectra (e.g. calibration spectra) where fewer propagation modes are populated \citep{Mahadevan2014b}. To combat this, RV systems use mechanical agitation to redistribute light between different modes and smooth over the interference pattern. Modern variants achieve sufficient mode mixing to average over the speckle patterns at the $<$10$^{-3}$ level.

\begin{marginnote}[]
\entry{mode}{A supported electromagnetic field distribution that can propagate through a waveguide.}
\end{marginnote}

\begin{marginnote}[]
\entry{modal noise}{spatially-variable intensity fluctuations in an optical fiber output due to interference of propagation modes.}
\end{marginnote}

Tracking the rate at which photons are being collected is also crucial for precise RV measurements. Knowledge of the exact location of the observatory relative to the solar system barycenter is needed to properly correct for the projected velocity of the observatory relative to the target. To achieve this, an accurate and precise flux-weighted midpoint time is needed \citep{WrightEastman2014}. In RV instruments this is done with a real-time \lq{}exposure meter\rq{} (EM) that records integrated light or low-resolution spectra alongside the science spectra to produce an intra-exposure flux time series. EMs in modern optical RV instruments pick off light upstream from the spectrometer focal plane through various techniques \citep[e.g. redirecting light that would otherwise overfill the spectrometer pupil, or collecting light from the 0th order of the spectrometer grating,][]{Landoni2014}.

\subsection{Spectrograph Architectures}\label{sec:architectures}

\begin{figure}[h]
\includegraphics[width=5in]{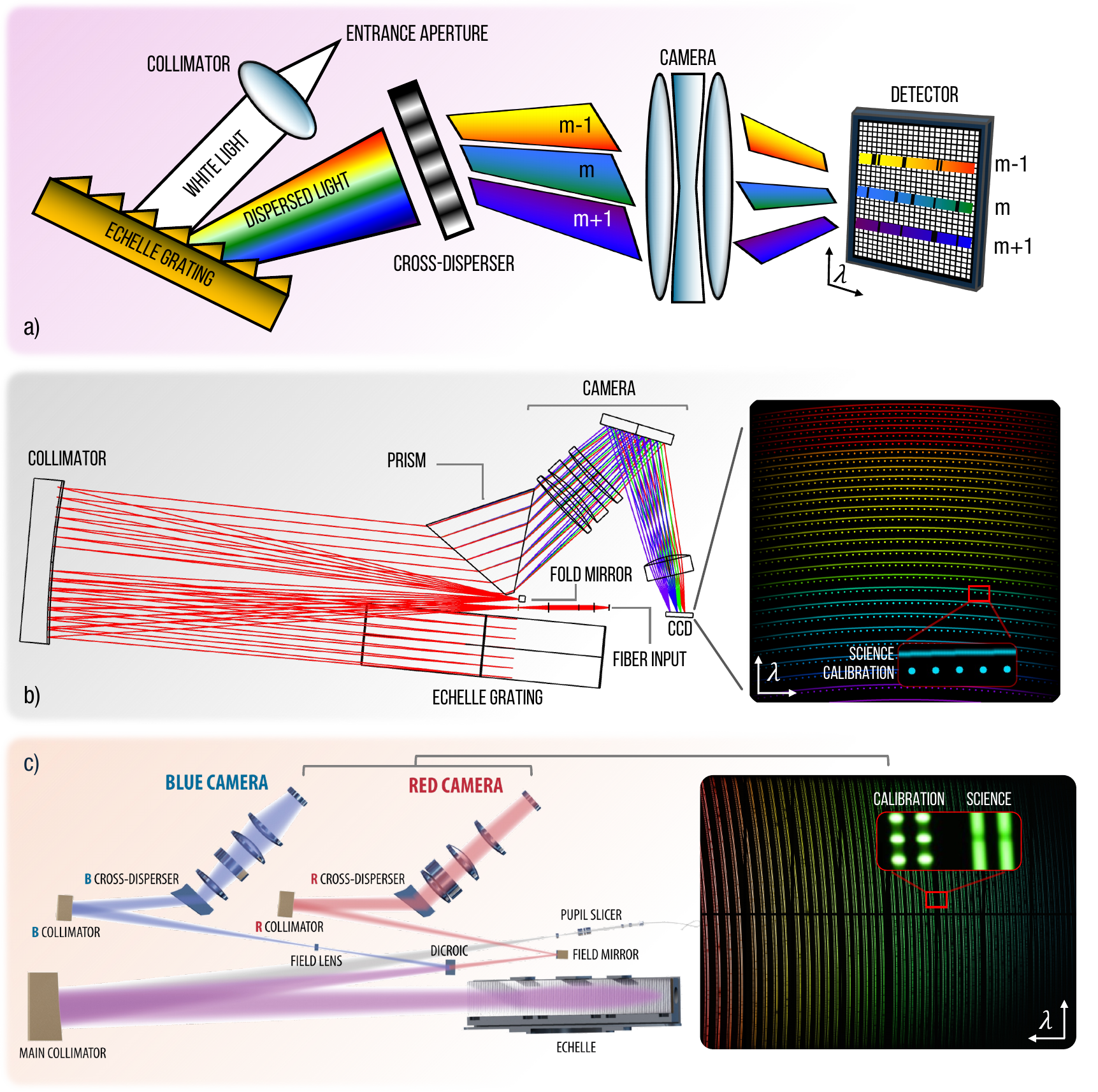}
\caption{a) Basic principle of a cross-dispersed, high resolution échelle spectrometer used for precise radial velocity (RV) measurements. Light from the entrance (generally a slit or optical fiber) is collimated and dispersed with a high blaze angle, low groove density, échelle grating. The overlapping échelle diffraction orders (m$\pm$n) are then separated using a second dispersive element, oriented with its dispersion direction at 90-degrees to that of the échelle grating.  This element is known as the ‘cross-disperser’ and is typically a prism or grating. The order-separated light is then imaged onto a focal plane using a set of camera optics. b) Optical layout of the NEID RV spectrometer \citep{Schwab2016}, which uses a symmetric \lq{}white-pupil\rq{} architecture \citep{Baranne1988} to record a broadband (380 - 930 nm) optical spectrum on a single large format CCD (right) using a prism cross disperser.  NEID leverages a simultaneous calibration fiber to track instrumental drift using a stabilized wavelength reference (right, inset). c) ESPRESSO optical layout \citep{Pepe:2021aa}, showing an example of an \textit{asymmetric} white-pupil design (beam is compressed after the primary collimator). ESPRESSO uses a pupil slicer to reformat the image of the input fiber into two separate spectral traces that are spatially offset (right, inset) and recorded using two separate cameras that are optimized for different wavelength regimes. This approach preserves high spectral resolution without significantly increasing the spectrograph pupil diameter.}
\label{fig:spectrometer_cartoon}
\end{figure}

The need to collect a wide spectral grasp while maintaining high resolving power motivates the use of large ($\sim$400-1200 mm long) échelle gratings paired with cross-dispersers. This combination produces a wide bandwidth, two-dimensional spectral format that can be recorded on a square or rectangular focal plane. An overview of a typical design is shown in Figure~\ref{fig:spectrometer_cartoon}, including example spectrometer optical layouts from existing instruments.

Multiple spectrometer design philosophies have been employed over the past decades for precise Doppler measurements, ranging from using uncontrolled or coarsely controlled systems that imprint truly simultaneous calibration spectra on top of the stellar spectrum \citep[e.g.][]{Vogt1994, Butler1996, Crane2008}, to environmentally-stabilized systems that minimize absolute motion of the recorded stellar spectrum at the {\ms}-level and rely on dedicated calibration channels to monitor and subtract residual instrument drift \citep[e.g.][]{Mayor2003, Jurgenson2016, Schwab2016, Gibson2024, Pepe:2021aa}. Many variants of this basic cross-dispersed design exist \citep{Vogt1994, Mayor2003, Crane2008, Spano2010, Schwab2016} that have been optimized for specific science goals and telescope architectures. The core of most modern RV spectrometer designs revolves around the \lq{}white pupil\rq{} approach \citep{Baranne1988} - a proven architecture for recording wide bandwidth, cross-dispersed échelle spectra at high resolution. This approach re-images the dispersed pupil onto the cross-disperser (see Figure~\ref{fig:spectrometer_cartoon} for an example), which can allow for better aberration control, smaller optics, and reasonable design flexibility.

More recently, optical \lq{}slicing\rq{} has been employed to simultaneously support high throughput and high spectral resolving power. This approach divides light incident on the spectrometer into a more conveniently-formatted illumination pattern that is narrower in the spectral dimension at the cost of being wider in the spatial dimension. Slicing is particularly crucial for larger telescopes which would otherwise have an untenable \'{e}tendue \citep{Seifert2012, Conconi2013, Gibson2024}, since the spectrograph beam diameter scales with telescope diameter at fixed spectral resolution. Slicing also plays a central role for instruments on smaller telescopes aiming to reach very high resolving power \citep{Jurgenson2016} and/or preserve high throughput \citep[e.g.][]{Quirrenbach2014}.

\begin{marginnote}[]
\entry{optical slicing}{reformatting of light in the pupil or image plane to a narrower image in the dispersion direction, which increases spectral resolution}
\end{marginnote}

Most RV instruments employ a dedicated calibration channel that is permanently fed with a wavelength reference source. This calibration channel (\lq{}CAL\rq{}) is spatially offset from the stellar spectrum channel (\lq{}SCI\rq). The inset images in Figure~\ref{fig:spectrometer_cartoon} show examples of this approach. This CAL channel provides an in-situ measurement of drift during observations, which can then be subtracted before the stellar RVs are computed. This technique requires that the differential motion across the entrance slit is minimal during an exposure such that the CAL channel \lq{}sees\rq{} the similar instabilities as the SCI channel. This assumption is reasonable for instruments that utilize thermally controlled or isolated vacuum chambers to minimize thermomechanical instabilities \citep{Mayor2003, Robertson2019, Stefansson2016, Gibson2024} and high performance optical fiber feeds to stabilize the illumination \citep[e.g.][]{Kanodia2023}. Instruments that follow this approach have routinely demonstrated $<$10 c{\ms} calibration error\footnote{defined as the residual signal between the SCI and CAL channels when both are illuminated with the same calibration source.}.

\subsection{Environmental Stabilization}\label{sec:stabilization}
All spectrographs have some level of thermal sensitivity. Refractive elements, typically used at the spectrometer entrance and in the camera(s), can have significant coefficients of thermal expansion (CTE) and thermo-optic coefficients (dn/dT). Changes in the temperature of these optics can produce significant motion of the spectrum at the focal plane and/or changes in the effective aberration distribution of the instrument PSF \cite[e.g.][]{Halverson2016}. Temperature fluctuations in the optical bench and mounts can also cause the relative optic positions to change. Both of these effects can be calibrated using reference sources, though not necessarily perfectly at the c{\ms}-level.

To maximize measurement sensitivity and alleviate performance requirements on both the wavelength calibration system and the data reduction pipeline most modern RV spectrographs are enclosed in actively and/or passively stabilized environments. These systems often employ a layered thermal control approach, stabilizing the surrounding environment with increasing precision moving from the ambient environment to the vacuum vessel to the spectrometer optical train. This general approach is used by multiple current RV facilities, including HARPS, ESPRESSO, CARMENES, NEID, and others. Material choice is a key design parameter when considering stability. Instruments have utilized a variety of low CTE materials for their optical benches/frames, such as stainless steel (HARPS, ESPRESSO), Invar (EXPRES, iLocater), and even Zerodur (KPF) to minimize thermal sensitivity.

Some instruments include active thermal control within the vacuum chamber to directly stabilize the temperature of the optical bench \citep[e.g. HPF, NEID,][]{Robertson2016, Stefansson2016}. In these systems, thermal conductivity is prioritized over low coefficient of thermal expansion. This results in fast response time and tight thermal control, at the cost of higher effective thermal sensitivity due to higher CTE materials being used.

\begin{figure}[h]
\includegraphics[width=5in]{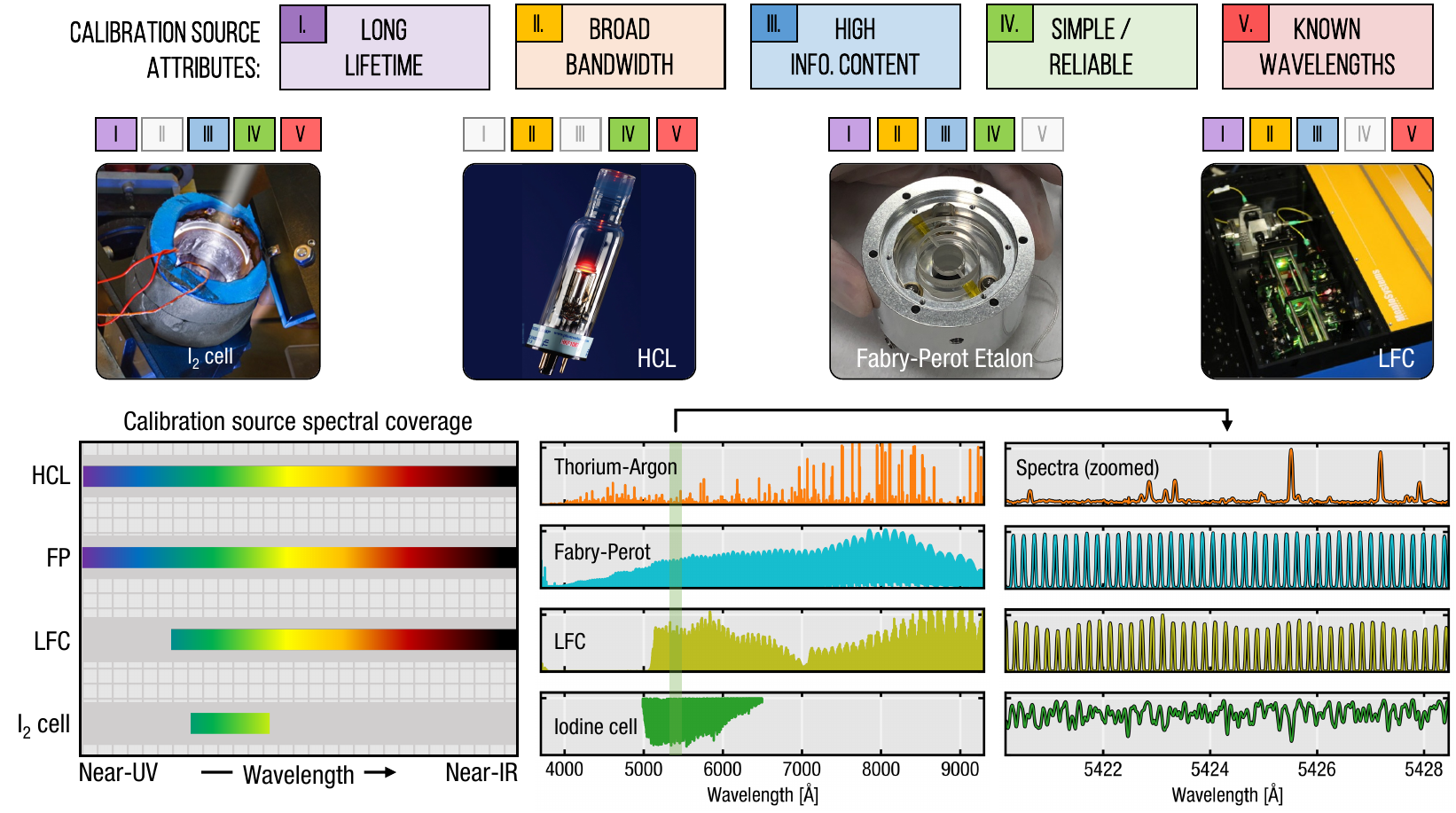}
\caption{Summary of modern optical wavelength calibration sources used for precise radial velocity measurements over the past decade, with approximate spectral coverage and general key attributes noted (top row). Essential features that makeup the ideal RV wavelength calibration source include long lifetime (I), wide spectral grasp (II), high density of spectral features, ideally uniformly bright and unblended and supporting of $<$10 c{\ms} calibration precision (III), low operational complexity and high reliability (IV), and intrinsic accuracy from spectral features that are intrinsically stable and traceable (V). Images of different calibration sources are shown in the middle row; individual calibration source descriptions are summarized in Section~\ref{sec:wavecal}. The bottom row shows approximate spectral coverage of each source (left), along with example recorded spectra from different RV facilities (middle), with a highlighted region of a few Angstrom (right). The zoomed spectra illustrate the differences in spectral uniformity between each source.}
\label{fig:cal_source_summary}
\end{figure}

\subsection{Wavelength Calibrators}\label{sec:wavecal}
Any motion of a recorded stellar spectrum due to the measurement system (as opposed to the star) must be removed using precise calibration sources. The underlying instrumental dispersion must be referenced against a known frequency standard to maximize measurement performance and repeatability. Typical calibration sources used to set and monitor the wavelength scale of RV instruments include molecular absorption cells, hollow cathode emission lamps,  laser frequency combs, and broadband Fabry–Pérot etalons. Each source provides different levels of calibration performance. A summary of key features for these calibrators is shown in Figure~\ref{fig:cal_source_summary}. Below we describe each source, focusing on calibrators used in the optical. 

\subsubsection{Iodine absorption cells}

Some of the first reliably precise RV measurements were made by integrating a temperature controlled cell of gaseous, molecular iodine (I$_{2}$) into the converging beam of the telescope \citep{Marcy1992, Butler1996}. The iodine acts as a transmission filter, imprinting thousands of narrow absorption lines in the 5000 - 6200 \AA\ region onto the stellar spectrum. The Iodine lines share a common optical path with the starlight, so any changes to the instrument profile will be reflected in the I$_{2}$ spectrum. The I$_{2}$ lines have precisely known, stable wavelengths \citep{Koch1984} and each cell is scanned using a Fourier Transform Spectrograph (FTS) to determine its individual transmission function. The star's instantaneous velocity is computed as a free parameter by modeling the observed spectrum as the instrument profile convolved with the product of the transmission function of the I$_{2}$ cell and the intrinsic stellar spectrum, including a Doppler shift. While this approach has multiple advantages, including truly common-path tracking of instrumental variations and simplicity of integration into existing high resolution spectrometers, the I$_{2}$ has a limited bandwidth ($\sim$1200 {\AA}) and as light must pass through the cell for the method to work there is a net efficiency drop in the system. Even with the high resolution template derived from the FTS spectra, thousands of free parameters are required to fully characterize the combined cell $\times$ stellar spectrum. For these reasons, I$_{2}$ cells are no longer used for the most precise RV measurements.

\subsubsection{Hollow cathode lamps}
Hollow cathode lamps (HCLs) consist of a heavy metal cathode filament encapsulated in a glass tube that is filled with an inert buffer gas (generally Argon or Neon). These lamps produce a dense set of emission features with well-characterized transition frequencies. The most popular HCLs used for RV calibration are Thorium-Argon (ThAr) in the optical \citep{Lovis-2007b, Redman:2014aa}, and Uranium-Neon (UNe) in the near-infrared (NIR) \citep{Redman:2011aa}. ThAr lamps in particular have enabled $\sim$50 c{\ms} calibration repeatability in stabilized systems over the past decade \citep{Dumusque:2018aa, Dumusque:2021aa}, and continue to calibrate many current RV instruments.

The non-uniform density and variable relative strengths of these features can lead to a significant number of line blends at the resolutions of typical astronomical spectrometers (see Figure~\ref{fig:cal_source_summary}). Additionally, the bright lines from the lighter inert gases used to fill typical HCL tubes introduce a significant number of variable parasitic emission features that can result in systematic calibration errors \citep{Lovis-2007b}. A careful curation of cathode spectral lines is thus required, which limits the precision of an HCL wavelength solution to $~\sim$ 20 c\ms\ \citep[Dumusque et al. 2025, submitted,][]{Dumusque:2021aa}. Another challenge with HCLs is that their spectral features drift at the level of a few \ms\ over the few year lifetime of the lamp (Dumusque et al. 2025, submitted), which can compromise long-term performance. Additionally, recent manufacturing issues have precluded the production of pure thorium filament lamps. Instead, these lamps are now fabricated using thorium-oxide filaments which introduces a wealth of contaminating transitions that are not well characterized, vary significantly from lamp to lamp \citep{Nave2018}, and reduce lifetimes.

\subsubsection{Laser Frequency Combs}
Laser Frequency Combs (LFC) are the current pinnacle of astronomical frequency standards, producing sharp emission features (modes) at wavelengths traceable to stabilized atomic transitions \citep{Murphy2007, Steinmetz2008, Metcalf2019}. LFC lines have typical spectral widths over 1000$\times$ narrower than the spectrometer resolution element. This maximizes calibration accuracy and also provides a reliable measurement of the instrumental LSF. LFCs used in astronomical applications are generally frequency stabilized by locking both the oﬀset frequency, $\nu_{\mathrm{offset}}$, and laser repetition rate, $\nu_{\mathrm{rep}}$, to a global positioning system-stabilized atomic clock, which can yield absolute precisions better than 10$^{-12}$ (equivalent to m{\ms} in velocity). The frequency of a given LFCs mode, $\nu_{\mathrm{n}}$, can be computed assuming both $\nu_{\mathrm{offset}}$ and $\nu_{\mathrm{rep}}$ are well known:

\begin{equation} \label{eq:laser}
    \nu_{n} = n\times\nu_{\mathrm{rep}} + \nu_{\mathrm{offset}}
\end{equation}

A core technology for current optical LFCs is a femtosecond pulsed mode-locked laser that produces a comb of evenly spaced optical frequencies \citep{Steinmetz2008}. These lasers have a native repetition rate that is too low  to be resolved by typical RV instruments (100s of MHz mode spacing, compared to $\sim$6 GHz spectral LSF\footnote{A 3 k{\ms} LSF width ($R$=100,000) corresponds to 6 GHz frequency width at 500 nm.}). These systems use a string of Fabry–Pérot filter cavities to filter the intrinsic mode spacing down to 10–30 GHz to maximize utility for spectrograph calibration. To generate the broad spectrum required to span an RV instrument's band pass, light produced by the mode-locked source is coupled into a non-linear broadening waveguide. Multiple stages of optical amplification are needed to reach the average power levels required to generate a broad spectrum in these systems.

Instead of using mode-locked lasers, Electro Optic (EO) combs use a continuous-wave laser with phase and amplitude modulators driven by stabilized radio frequency sources to produce a broad LFC spectrum. This typically makes EO systems more compact, robust, and field-deployable than mode-locked femtosecond systems. EO comb modes can be natively generated at 10-30 GHz mode spacings using commercially available components, making their design simpler than their mode-locked counterparts. Because of the nature of the comb generation, EO combs are also relatively simple to tune in both offset frequency (set by the CW laser frequency) and mode spacing (set by the modulator frequencies). EO combs have only been developed reliably in the NIR for RV applications \citep[e.g][]{Metcalf2019} due to the abundance of optoelectronic infrastructure available (and lack thereof in the optical), though efforts are ongoing to adapt this approach to bluer wavelengths.

Regardless of the style of system, LFCs generally have a high degree of engineering and operational complexity compared to other calibration sources, requiring many photonic and electronic subsystems to work in concert for prolonged, uninterrupted operation. This remains an outstanding challenge with these technologies, particularly in the optical where generation of the requisite supercontinuum is difficult. In commercial mode-locked systems, like those used at ESPRESSO, NEID, KPF, EXPRES, and HARPS, tailored photonic crystal fibers (PCFs) are used to nonlinearly broaden the LFC spectrum. To achieve a broad supercontinuum, the average power injected into the PCF is high (typically multiple watts, on average). However, these PCFs are prone to damage under high peak powers from femtosecond pulses and the nonlinear process degrades the fibers over time by physically altering the structure of the fiber face. This can compromise long term reliability and increases operational cost and complexity costs significantly.

\begin{marginnote}[]
\entry{supercontinuum}{a broad spectrum of light generated by sending a pulsed laser signal through a nonlinear optical medium.} 
\end{marginnote}

\subsubsection{Fabry–Pérot Etalons} \label{sec:FP}
Fabry–Pérot etalons (FPs) are interferometric cavities that filter incident broadband light, transmitting only those wavelengths that satisfy specific interference conditions. These conditions depend on the cavity spacing/geometry and the coatings applied to the mirrors. The FP transmission function is a comb-like spectrum that is rich in calibration features over wide wavelength ranges. The cavity's free spectral range ($\Delta\nu$, the frequency spacing between spectral interferometric modes) can be parameterized to first order by the refractive index, $n$, and the effective cavity length, $L$:

\begin{equation} \label{eq:fp_spacing}
    \Delta\nu = \frac{c}{2nL}
\end{equation}

The etalon resonance frequencies, $\nu_n$, follow the general relation below, assuming perfect knowledge of the mode number $m$, the absolute cavity mode number{$\nu_o$}, and $\Delta\nu$):

\begin{equation}
    \nu_n = \nu_o + m\times\Delta\nu
\end{equation}

Because of the simplicity and flexible design parameters (cavity length and coatings can be easily customized\footnote{The cavity spacing and mirror reflectivity set the sharpness of the etalon trannsmission function.}), FPs are used at numerous RV facilities for calibration. To avoid introducing instabilities into the FP cavity, these systems must be environmentally controlled and illuminated in a stable, repeatable way. Illumination stability is critical, as this directly affects the effective cavity phase. As such RV FP systems are fed with an optical fiber. A broadband illumination source, typically a broadband lamp or supercontinuum source, is used to illuminate the FP. Many iterations of FPs have been explored, from fiber-based \citep[e.g.][]{Halverson2014} to vacuum cavities using both multi mode \citep{Wildi2010, Cersullo2017} and single mode \citep{Sturmer2017} fiber injection. Current systems that use FPs for tracking relative instrumental drift have demonstrated c{\ms}-level calibratability over short timescales \citep[e.g.][]{Gibson2024, Pepe:2021aa}

A drawback in FPs is that the line frequencies are not know apriori. The absolute mode number is dependent on the exact cavity phase delay, which is not possible to characterize without additional calibrations. The drift rate of the FP cavities has also been shown to be chromatic \citep{Terrien2021, Kreider2025}, potentially complicating their use for the highest precision drift correction. For these reasons an external absolute reference (e.g. an LFC) is required to re-anchor the FP spectrum periodically.

\subsection{Detectors}\label{sec:detectors}

Large-format, highly-efficient detectors are required to record the wide swaths of stellar spectrum needed for precise RV measurements. In the optical, high performance charge-coupled devices (CCDs) are used to image the 2D stellar spectra produced by the spectrograph camera(s). Modern CCDs have high quantum efficiency (QE) and can be fabricated into large arrays (up to 10k $\times$ 10k pixels per device). CCD performance has been steadily improving over the past decades of astronomical use, with QEs exceeding 90\% over many hundreds of nanometers of wavelength with broadband anti-reflection coatings applied.

\begin{marginnote}[]
\entry{quantum efficiency}{the effective efficiency of conversion between incident photons and recorded photoelectrons in a detector pixel.} 
\end{marginnote}

While highly efficient, these devices suffer from a variety of noise sources. These sources include read noise and dark current, which are properties of both the active silicon layer and readout electronics, and a variety of systematic errors associated with the charge transfer and readout process. Importantly, these noise sources span both random and systematic behavior. Read noise and dark current are unavoidable contributors and are functions of detector temperature, applied readout waveforms, readout electronics, and electrical environment. The largest CCDs are fabricated using a multi-step lithographic writing process. These discrete steps can introduce periodic errors in the pixel response that introduce discontinuities in the otherwise smooth wavelength solution \citep{Coffinet2019}. The efficiency of charge shuffling between pixels during readout is also imperfect and signal to noise dependent \citep{Blake2017}. This leads to a systematic \lq{}trail\rq{} of leftover charge in the readout direction that can bias the RV measurements \citep[][]{Bouchy2009, Halverson2016, Blake2017}. The pixel potential well structure is not static as a function of signal strength, which leads to the so-called brighter-fatter effect \citep{Antilogus2014, Guyonnet2015}. Finally, The silicon active layer of CCDs can act as an interferometric cavity, especially for longer wavelengths, leading to spatial fringing in the recorded flux distribution that can be difficult to divide out reliably.

\begin{marginnote}[]
\entry{fringing}{interference patterns due to internal reflection within the detector's silicon layers}
\end{marginnote}

Standard CCDs are impractical for use in the NIR, as the band gap of silicon does not support wavelengths red-ward of $\sim$1 micron. Instead, Complementary Metal-Oxide-Semiconductor (CMOS) arrays are used. NIR CMOS arrays are fabricated with more exotic material combinations, such as mercury cadmium telluride (HgCdTe), to maximize sensitivity at NIR wavelengths. CMOS arrays have individually addressable pixels with integrated readout circuits, which allows for rapid readout and non-destructive measurements of the accumulated charge. Non-destructive reads allow for sampling of the flux rate multiple times during an exposure, which allows for improved signal estimation. This sequence of non-destructive reads also provides the intra-exposure flux information needed to properly compute the flux-weighted exposure midpoint without the need for an external exposure meter. While superior from a flexibility standpoint, NIR CMOS arrays suffer from a variety of noise sources. These include variable gain (each pixel is essentially a self-contained amplifier), interpixel capacitance (coupling of charge between adjacent pixels), image persistence (leftover charge after readout), and sub-pixel QE variations \citep{Ninan2019} due to lattice defects (which manifest as cross-hatch patterns), to name a few. \citet{Bechter2018} and \citet{Artigau:2018aa} provide comprehensive overviews of the challenges of NIR HgCdTe CMOS arrays for precise RV spectroscopy.

\subsection{The Future} \label{future:instrumentation}

Recent advancements in adaptive optics (AO) systems have enabled the use of single-mode fibers (SMF) to deliver light to downstream spectrometers \citep[e.g.][]{Jovanovic2017, Bechter2020b}. With AO correction approaching the theoretical limits in the NIR, $\sim$10's of percent of light can be coupled into an SMF. This is comparable to the coupling efficiency of seeing-limited RV systems in the optical (typically $\sim$50\%). Use of an SMF eliminates classical intensity mode interference patterns observed in multi-mode fibers (see above sections), though there remains some variability due to multiple supported polarization modes \citep{Halverson2015b, Bechter2020a, Gibson2025}. The output mode of an SMF is fundamentally stable spatially, which ensures the spectrometer is presented with a perfectly repeatable intensity pattern in both the near and far fields. Importantly, a spectrometer fed by an SMF can be designed at the diffraction limit, resulting in a significantly smaller optical system when compared to their seeing-limited counterparts for the same spectral resolution \citep{Schwab2012}. Multiple diffraction-limited RV systems that aim to reach $<$50 c{\ms} in the NIR are currently in development \citep[e.g.][]{Cale2023, Konopacky2023, Crass2022}, though pushing into optical wavelengths will require significant improvements in AO performance. 

The RV calibration \lq{}dream-machine\rq{} has yet to be developed though the community has converged on a set of ideal properties (see Figure~\ref{fig:cal_source_summary}). Such a system would ideally have a dense forest of regularly-spaced emission features, each of which is traceable to a stable atomic transition, span 100's of nm of spectral coverage, have uniform intensity, minimal consumables, and 100\% reliability. LFCs remain the most tantalizing technology for this and multiple prospective technologies are being developed to address these needs. New waveguide technologies could extend the reliability and simplicity of EOM LFCs to optical wavelengths, which could alleviate the need for spectrally-filtered mode-locked systems for supercontinuum generation \citep{Cheng2024}. Comb generation using microresonator cavities has also been explored for RV spectroscopy in recent years \citep{Suh2019}. These \lq{}microcombs\rq{} leverage multiple nonlinear effects in whispering-gallery microresonators, typically made of materials like MgF$_\mathrm{2}$ or SiN, though there has yet to be a field-deployable system developed for RV spectroscopy.

Reliably reaching the requisite instrumental measurement capability to detect Earth-like planets may require a host of new technologies, especially if the community embraces a next generation EPRV system that is to be replicated for multiple observatories to maximize survey efficiency \citep[e.g.][]{Luhn2023}. Numerous advancements in photonic technologies for astronomical spectroscopy are being explored for RV measurements, from mode-separating waveguides (photonic lanterns) to dispersive elements (array waveguides). For a comprehensive review of RV astrophotonic technologies, see \cite{Minardi2021}. Improvements in detector technologies, such as the advent of large-format optical CMOS arrays, and grating fabrication methods (e.g. lithographically etched grooved surfaces) also show great promise for enabling both higher performance and greater efficiency.

\section{DATA REDUCTION PIPELINE AND RV COMPUTATION} 
An extremely stable instrument operating at the highest possible spectral resolution is key for EPRV measurements. However, one cannot neglect the work required to develop and maintain the instrument's data reduction pipeline (DRP) which must extract the spectral and wavelength calibration information from raw detector images with the highest possible fidelity. Here we use fidelity to mean that the pipeline should produce a stellar spectrum that accurately reflects the starlight as it entered the telescope at the time of measurement, having been corrected for all known telescope, instrumental, and detector systematics. In this section we omit the extraction and calibration methods for I$_{2}$ cell instruments, focusing instead on reviewing current state-of-the-art techniques for stabilized RV spectrographs.

\subsection{Spectral Extraction: From Raw Images to Wavelength Calibrated Spectra}

\begin{figure}[h]
\includegraphics[width=5in]{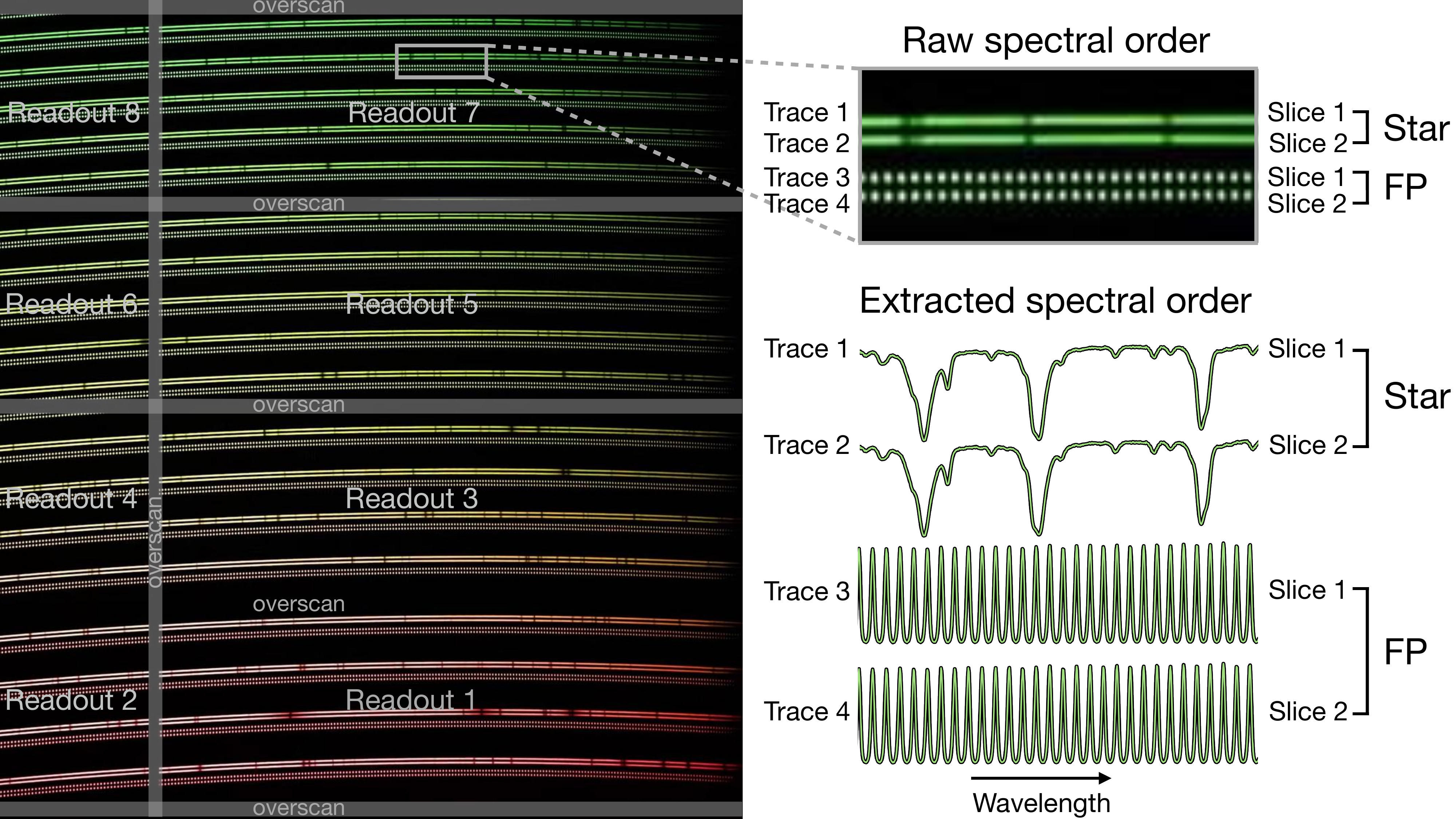}
\caption{\emph{Left:} Zoom on a raw image obtained by the red detector of the ESPRESSO spectrograph, for a stellar observation with simultaneous FP calibration. We can see 8 readouts out of the 16 of the detector, each surrounded by overscan regions used for bias measurement. \emph{Top right:} Zoom on a spectral order which contains four traces, two corresponding to the sliced stellar light, and two for the sliced FP light. \emph{Bottom right:} After extraction and wavelength calibration we obtain, for each trace of a spectral order, a 1D spectrum as a function of wavelength.}
\label{fig:spectral_format}
\end{figure}

As seen in Sect.~\ref{sec:architectures}, high spectral resolution is obtained by looking at high diffraction orders and separating them using a cross-disperser. The resulting spectral format on the CCD will consist of consecutive diffraction orders that together cover the entire spectral range of the spectrograph (Figure \ref{fig:spectral_format}). Each spectral order can also have multiple \lq{}traces\rq{}, corresponding to different light-paths injected into the spectrograph. These traces can capture different slices of an incoming fiber (to increase spectral resolution without losing light) or different fibers used to record the nearby sky or light from a calibration lamp alongside the target.

The goal of the spectral extraction process is to begin from the raw detector images and derive for each trace across all spectral orders a wavelength-calibrated, 1D spectrum that has been corrected for instrumental systematics. We note that our description below does not delve into the specific algorithms used for processing calibration and science frames, but more detail can be found in various instrument DRP resources such as the ESPRESSO DRP \citep{ESPRESSO-Manual} and NEID DRP (https://neid.ipac.caltech.edu/docs/NEID-DRP).

\subsubsection{Detector specific properties and systematics corrections}

As described in Sect.~\ref{sec:wavecal}, visible and NIR spectrographs do not use the same detector technology and so the detector systematics differ between these two families of instruments and require different treatments. We focus below on visible spectrographs, which currently demonstrate the best RV precision, and thus describe how to handle common CCD-centric systematics. We refer readers interested in the nuances of NIR detectors to \citet{Artigau:2018aa} for a helpful overview of H4RG NIR CMOS detector characterization.

In order to extract the flux from different traces and correct for detector systematics, a series of specific calibration frames are required. Bias calibrations are needed to measure the detector's read-out-noise (RON), the detector's amplifiers' mean bias and their residual. Dark calibrations are required to measure the pixel dark-current and to identify \lq{}hot\rq{} pixels that present excessive dark-current. 2D flat-field calibrations are used to measure the gain of the detector and detect \lq{}bad\rq{} pixel that do not present a linear response to light. Finally, flat-field calibrations are required to measure the pixel-to-pixel sensitivity. A more detailed description of these standard CCD calibrations can be found in Section~A of the Appendix. 

\begin{marginnote}
\entry{gain}{the inverse of the photoelectron to analogue-to-digital (ADU) conversion factor}
\end{marginnote}

\begin{marginnote}[]
\entry{trace profile}{the normalized flux distribution in cross dispersion of a trace for all positions in spectral direction}
\end{marginnote}

Once the above-mentioned calibration frames and derived products are obtained, the information required to properly extract the flux of each individual trace from each spectral order is available. Master bias residual frames are used to remove remaining systematics once the bias measured from overscan regions is subtracted. The gain is used to transform analog-to-digital (ADU) units into a number of photo-electrons. Measurements of the gain, RON and dark current are critical to properly propagate errors during extraction. Bad and hot pixel maps are used to reject spurious pixels when extracting the signal. Finally, master flat-field calibrations are used to localize the spectral traces, extract their corresponding flux to measure pixel-to-pixel variability, and obtain the trace profile in cross-dispersion. While the bias residual correction and the propagation of the RON and dark current are not critical for high S/N observations, they must be performed when looking at very faint objects.

\subsubsection{Extracting the spectra} \label{spectral_extract}

Fiber-fed EPRV spectrographs produce spectral traces with a relatively symmetric slit image on the detector perpendicular to the dispersion direction. In other words, one can assume (though not entirely correctly) that all flux from a trace along a detector column corresponds to the same wavelength. Therefore, the flux of each trace can be derived by summing the flux in this cross-dispersion direction, using a window of fixed width centered on the central position of each trace. Adding pixels with little to no flux will lower the S/N by increasing the RON and dark current noise, and so a weighted sum taking into account the relative flux of each pixel must be performed. This strategy, known as optimal extraction and first published in \citet{Horne:1986aa}, requires the user to know the trace profile at each position in dispersion to perform the weighted sum. For EPRV spectrographs, this trace profile is measured from high-S/N master flat-field calibrations.

An optimally-extracted spectrum will still be affected by pixel-to-pixel variations and fringing effects. To correct for these systematics, we extract the flux of the master flat-field calibration frame in exactly the same way. The output spectrum will present the pixel-to-pixel variations, in addition to slow flux variations due to the blaze function of the spectrograph and the calibration lamp temperature. The latter can be corrected for as the spectral energy distribution of the lamp follows a black body at the lamp's temperature. Then a high- and low-pass filter can be used to separate the blaze function contribution from the pixel-to-pixel variation. Dividing an extracted spectrum by this latter product will correct for pixel-to-pixel variations and fringing. In this approach the spectrum is extracted using pixel-to-pixel variations, fringing, and a trace profile all measured using a master flat-field calibration that is often taken several hours apart from the science observations. This is only possible because EPRV spectrographs are extremely stable on a daily timescale. 

\begin{marginnote}[]
\entry{blaze function}{smooth, wavelength-dependent intensity profile of an échelle order caused by varying diffraction efficiency and general spectrograph design.}
\end{marginnote}

Rather than measuring the pixel-to-pixel variation and the profile on the master flat-field calibration before using them for correction, it is also possible to directly extract the trace of interest using the master flat-field calibration without deriving the intermediate products. This method, known as flat-relative optimal extraction \citep{Zechmeister:2014aa}, presents the advantage of performing optimal extraction without requiring the trace profile and while simultaneously correcting for pixel-to-pixel variation and fringing.

\subsubsection{Wavelength solution} \label{sec:wavesol}

Once the spectrum is extracted, the actual information obtained for each trace is its flux as a function of pixel position along the dispersion direction. Since optimal extraction assumes that each extracted pixel corresponds to a specific wavelength, a wavelength solution is then required to map pixel position to wavelength. To do so, we need to obtain a calibration in which a spectrum presenting sharp spectral features at well quantified wavelength locations is recorded on the different traces of the spectrograph. EPRV spectrographs use spectra from either a HCL or LFC to generate their wavelength solutions (Sect.~\ref{sec:wavecal}). For HCLs, multiple line tables listing the  spectral features and their respective wavelengths exist and should be used as guides: \cite{Sharma:2021aa} and \cite{Sarmiento:2018aa} for Uranium and Dumusque et al. 2025 (submitted), \cite{Dumusque:2021aa}, \cite{Redman:2014aa} and \cite{Redman:2011aa} for Thorium. For LFCs, equation~\ref{eq:laser} can be used to derive the wavelength of all emission lines.

A precise wavelength solution can be derived in several ways. The simplest is to fit an independent wavelength solution for each spectral order by first performing an analytic fit (typically a Gaussian or polynomial) to determine the central pixel position of each HCL or LFC feature and then interpolating between those central pixels using polynomials or other smooth functions \citep[e.g., a cubic Hermite interpolating polynomial,][]{Zhao:2021aa} to derive the wavelength across the entire order. This works well for LFC spectra, which contain hundreds of high S/N spectral lines per spectral order, but is more challenging for HCLs which generate far fewer spectral features that vary strongly in intensity. As demonstrated in \cite{Dumusque:2021aa} and \citet{Dumusque:2018aa}, the sparsity of Thorium lines (equivalent for Uranium lines) per spectral order, coupled with low flux on the edges of each order due to the underlying blaze function, makes a polynomial fit unconstrained on the edges. This induces instabilities in the global wavelength solution of up to 80 c\ms\ RMS on HARPS-N. One solution to this problem is to use the spectrum from an FP in combination with the HCL spectrum. The FP spectrum presents a density of lines similar to an LFC, for which the separation between lines depends on the length and the refractive index of the FP cavity (Equation~\ref{eq:fp_spacing}). The spectral lines in the HCL spectrum can associate wavelengths to adjacent FP peaks and then the FP dispersion relation can be used to assign a wavelength to each FP spectral line \citep[e.g.][]{Cersullo:2019aa,Bauer:2015aa}. 

It is also possible to perform a wavelength solution based only on the HCL by simultaneously modeling all the orders, thereby reducing the instabilities that come from the sparsity of HCL spectral lines. As instrumental drift induces smooth variations over the detector with time, a solution is to model this drift with respect to a reference, by using a 2D polynomial fit over the entire detector. Then, correcting the reference wavelength solution using this smooth drift gives us the desired instantaneous wavelength solution. For HARPS-N such a method has been shown to produce wavelength solutions that are as stable as when combining the spectra of an HCL and an FP \citep{Dumusque:2021aa}. 

We warn that observers should not contemporaneously record the science spectrum on one trace and the spectrum of an HCL or an LFC on another trace to provide a simultaneous wavelength solution, as each trace will have a different LSF \citep[e.g.,][]{Schmidt2021}. Instead, the wavelength solution for a science observation should always be obtained using the same trace from which the science spectrum was extracted. To address the potential for instrumental drift between the acquisition of the wavelength solution calibration and the science observation, observers can illuminate a second trace (not the one used for the science and wavelength calibration spectra) with an FP spectrum during both the wavelength calibration and the science observation. As FPs drift $<$10 c\ms\ per day, the relative drift of the FP between the calibration and the science observations can be used to correct the wavelength solution from the instrument drift.

\subsection{RV Measurements and Other Post-processing Outputs}

Once an échelle-order, flat-field corrected, wavelength-calibrated spectrum is extracted, numerous products can be derived for precise RV work including the RV of the stellar spectrum and a suite of activity indices. It is also possible to correct for systematics that are challenging to model at the extraction level.

\subsubsection{Post-processing} 

The wavelength of an extracted spectrum will be in the Earth's reference frame, at the location of the telescope. To measure the RV of the star with respect the Solar System Barycenter (SBB), the Earth's velocity at the telescope location, projected along the line-of-sight of the observed object, has to be corrected for. This velocity, known as the barycentric Earth RV (BERV), is mostly affected by the Earth's orbit around the Sun ($\sim$30 \kms) and rotation ($\sim$500 \ms) as its precession, nutation, and variable rotation contribute only at the c\ms\ level. Most modern RV data pipelines use the BARYCORR package which provides corrections at the 1 c\ms\ level \citep[][]{WrightEastman2014}. 

Observations at low S/N can be contaminated by moon light, which imprints the solar spectrum on top of the stellar spectrum. If the sky spectrum is recorded contemporaneously on a different fiber of the spectrograph, moon-light contamination can be mitigated by subtracting the sky spectrum from the stellar spectrum if the relative efficiency between fibers is known \citep[e.g.][]{Roy2020}.

RV spectra are contaminated by telluric spectral lines formed in the atmosphere of the Earth. While in the visible we can exclude strongly contaminated spectral regions without significantly impacting RV precision \citep[e.g.][]{Artigau:2014aa,Cunha:2014aa}, in the NIR correction is mandatory to reach the 1 \ms\ precision. Tools like \emph{Molecfit} \citep[][]{Smette:2015aa}, TAPAS \citep[][]{Lallement:2025aa} or the strategy discussed in \cite{Allart:2022aa} and implemented in the ESPRESSO DRP can be used to model and correct for tellurics.

Time conveys significant information regarding instrumental systematics and a time series of spectra from the same object can be used to probe and correct for the behavior of the instrument with time. By analyzing spectral-time series residuals after removing an average stellar spectrum one can identify and remove systematics that are challenging to address at the extraction level, such as ghost contamination, cosmic rays, and interference patterns. This is the approach of the YARARA framework \citep[][]{Cretignier:2021aa} which has demonstrated a 20\% increase in RV precision when post-processing HARPS data. 

\subsubsection{RV measurement} 


The most popular method to compute precise stellar RVs is the cross-correlation technique \citep[e.g.][]{Griffin1967,Baranne-1996}. By cross-correlating the stellar spectrum with a binary template whose values correspond to the line RV content \citep[][]{Pepe-2002a,Lovis:2010aa} at the central wavelength of each spectral line and zeros elsewhere, we obtain a high S/N mean line profile called the cross-correlation function (CCF). Fitting the CCF with a Gaussian including a continuum provides not only the RV, but also the contrast (i.e. amplitude), the full width at half maximum (FWHM) and the mean of the profile. The mean of the fitted Gaussian is a direct measure of the star's radial velocity relative to the template. Although the template must resemble the spectrum of the observed star, a perfect match is not required and so a few carefully curated reference templates can be used to cover late-F to early-M dwarfs\footnote{The templates designed for ESPRESSO are available at \url{https://ftp.eso.org/pub/dfs/pipelines/instruments/espresso/espdr-kit-3.3.12-1.tar.gz}, with individual templates located in espdr-calib-3.3.12/cal/ESPRESSO\_[ST].fits where ST corresponds to different spectral types.}. To obtain a global RV that is not affected by color variations of the spectrum (which can occur for observations taken in poor conditions or at high airmass) the weight of each order in the weighted sum to obtain the global RV should be fixed and not allowed to vary with time. 

For M dwarfs, the density of spectral lines becomes so high that using a template with a finite number of spectral line positions will prevent us from extracting the maximum RV content of the spectrum. Instead, we use a template matching approach in which a high S/N template is constructed using all available stellar observations and an RV shift is then computed between each individual stellar spectrum and the template using a least-squares optimization \citep[][]{Anglada-Escude-2012, Zechmeister:2018aa} or a Bayesian approach \citep[][]{Silva:2022aa}. Template matching leverages the entire spectral content to compute an RV and, therefore, an optimal extraction of the RV is achieved. 

Another RV measurement method known as line-by-line (LBL) performs template matching for each individual spectral line and derives a corresponding RV using the method described in \citep[][]{Bouchy2001}. A global RV of the analyzed spectrum can be obtained by performing a photon-noise weighted average of all line RVs \citep[][]{Dumusque:2018aa,Artigau:2022aa}. LBL allows the user to detect spectral lines strongly contaminated by tellurics, detector systematics, or other unknown effects, as such lines will deviate from the main RV statistics. It is possible to strongly down weight those lines in the final weighted average, producing a more precise RV for NIR spectra \citep{Artigau:2022aa}. \cite{Dumusque:2018aa} and \cite{Lafarga:2023aa} demonstrated that LBL can also be used to identify lines that are particularly sensitive or insensitive to magnetic activity. 

We note that template matching and LBL require a template built from at least a dozen observations of the same star so that the template is at a much higher S/N than the individual observations and does not introduce noise into the RV derivation. In addition, those spectra have to cover a wide range in BERV so that when combining them in the stellar rest frame, unwanted signals fixed in the Earth rest-frame, such as telluric lines or detector systematics, are mitigated \citep[][]{Silva:2025aa}. For these reasons, while the CCF technique is the most robust method to derive RVs the LBL technique is much more effective when observing in the NIR or at high S/N to reach the best possible RV precision.

\subsubsection{Activity indices} \label{act_index}

The photosphere of a star is not static in time, and several different physical processes linked to convection and magnetic fields will induce changes in spectral line shapes, which will impact the derived RVs (Section~\ref{stell_act}). The most commonly used RV activity indices are the S-index and the related log(R'$_{HK}$) \citep[e.g.][]{Wilson-1968,Noyes-1984}, based on the chromospheric emission in the core of the CaII H and K lines at 3934 and 3969 \AA. See \cite{Maldonado:2019aa} and references therein for more activity indexes derived from individual spectral lines. The sun's unsigned magnetic flux, $|\hat{B}_{obs}|$, is also strongly correlated with its RV time series \citep{Haywood:2022aa}. While this quantity cannot be measured for other stars, \cite{Lienhard:2023aa} demonstrated that the Zeeman broadening of spectral lines from unpolarized spectra provides an indirect measure of $|\hat{B}_{obs}|$.

Activity indexes sensitive to magnetic activity in the NIR are not as well known, but a recent study from \cite{GomesdaSilva:2025aa} identified 17 lines in the NIR that can be used to detect rotational periods. It has also been shown in the NIR that line depth varies significantly with magnetic activity due to effective temperature variations induced by active regions. Therefore, it is possible to measure the differential signal in temperature over time, which provides a robust magnetic activity proxy \citep[][]{Artigau:2024aa}.

The Gaussian fit to the CCF of a given spectrum can also be used to investigate the star's activity. The FWHM, contrast, and CCF bisector are all sensitive to line-shape variations induced by stellar variability but not to pure Doppler shifts \citep[see e.g.,][]{Queloz-2001, Figueira-2013, Cegla:2019aa}. Comparison with these indices can therefore demonstrate whether or not significant signals in RVs have a variability counterpart, helping to differentiate planet-induced variations from other perturbing stellar systematics.

\begin{marginnote}[]
\entry{CCF bisector}{curve connecting the midpoints between the two wings of a CCF}
\end{marginnote}

\begin{marginnote}[]
\entry{CCF BIS SPAN}{The difference between the value at the top and bottom of the CCF bisector curve \citep[][]{Queloz-2001}.}
\end{marginnote}

\subsection{The Future} \label{future:extraction}

The solution adopted for mitigating instrumental drifts over time, to enable long-term RV precision, has been to stabilize spectrographs to the extreme. Yet this still leaves hundreds of \ms\ of absolute variation over years due to mechanical stress and optical aging \citep[e.g.][]{Dumusque:2021aa}. To reach the sub-\ms\ level, wavelength solution calibrations are used to reset the RV zero point of the spectrograph on a daily basis. However, the measured drift of the spectrograph is not a perfect Doppler shift, but rather a change in the instrument's 2D PSF over time. Because the measured spectrum is a convolution of the initial spectrum with that PSF, an asymmetric variation of the 2D PSF over time induces a different effect on the spectrum of a HCL or a LFC, with unresolved lines (Sect.~\ref{sec:wavecal}), than on a spectrum from a star or a FP, which both present resolved lines. The asymmetric 2D PSF variation as a function of time induces a divergence between the wavelength calibrations and observations, changing the shape of spectral lines which decreases the spectral fidelity and induces a trend in the RVs \citep[][]{LoCurto:2015aa,Schmidt2021}. 

A first solution to this problem is to model the 1D LSF and de-convolve it (or forward model the spectrum) to obtain the spectrum without 1D LSF contamination. This should be done for the wavelength solution calibrations, but also for the stellar spectra to obtain high fidelity. Some preliminary studies have shown that this is possible at the level of extracted frames \citep[][]{Hirano:2020aa} and it has been tested at the sub-\ms\ level on a few ESPRESSO calibrations to demonstrate wavelength solution accuracy \citep[][]{Schmidt2024}. Establishing the utility of 1D LSF modeling for EPRV work will require successful application of this technique to time series data, which has not yet been done.

Working at the level of the extracted spectra simplifies the problem. However, optimal extraction as described in Sect.~\ref{spectral_extract} is only optimal if the 2D PSF is a separable function of the two directions defined by the detector grid of pixels. If not (which is always the case to some extent due to varying slit tilt, correlation between neighboring cross-section profiles, and strongly varying flux levels in the spectral lines) then optimal extraction will degrade the resolution and S/N of the extracted spectrum. In addition, the trace profile needed for optimal extraction varies strongly between the spectrum of a continuous light source and a spectrum with sharp emission lines, such as a stellar spectrum or wavelength calibration source. Using the profile measured on the master flat-field for extracting the spectrum of a star or a wavelength solution calibration is thus not optimal, as demonstrated in \cite[][]{Schmidt2021}. To reach the highest possible spectra fidelity and long-term RV precision, point-spread function modeling and extraction should be done on the 2D raw frames, as proposed by the \lq{}spectro-perfectionism\rq{} framework \citep{Bolton:2010aa}. Although perfect in theory, such an extraction is non-trivial to apply to real data, is computationally expensive, and the only application of it to precise RV work thus far did not demonstrated any improvement over optimal extraction \citep[][]{Cornachione:2019aa}. However, we note that this comparison was performed on MINERVA, which is not an EPRV spectrograph, and the 2D PSF was modeled on sparse hollow cathode lamp emission lines. The output spectrum from an LFC should be used to obtain a much more precise 2D PSF model.

Time series of extracted spectra, both from calibrations and stellar observations, convey substantial information regarding instrumental signal residuals. The YARARA framework has demonstrated that instrumental signals can be probed at the level of extracted stellar spectra and that correcting for them at this level before measuring an RV improves RV precision by 20\% \citep[][]{Cretignier:2021aa}. However, YARARA only works with merged extracted spectra as input, which mix some instrumental signal information happening at the 2D level on the raw frame. A similar framework should be extended to \'echelle order extracted spectra, and not only to stellar spectra, but also calibrations.

\section{MITIGATING STELLAR VARIABILITY} \label{mit_stell_act}

\subsection{Stellar Variability Overview} \label{stell_act}

\begin{figure}[h]
\includegraphics[width=5in]{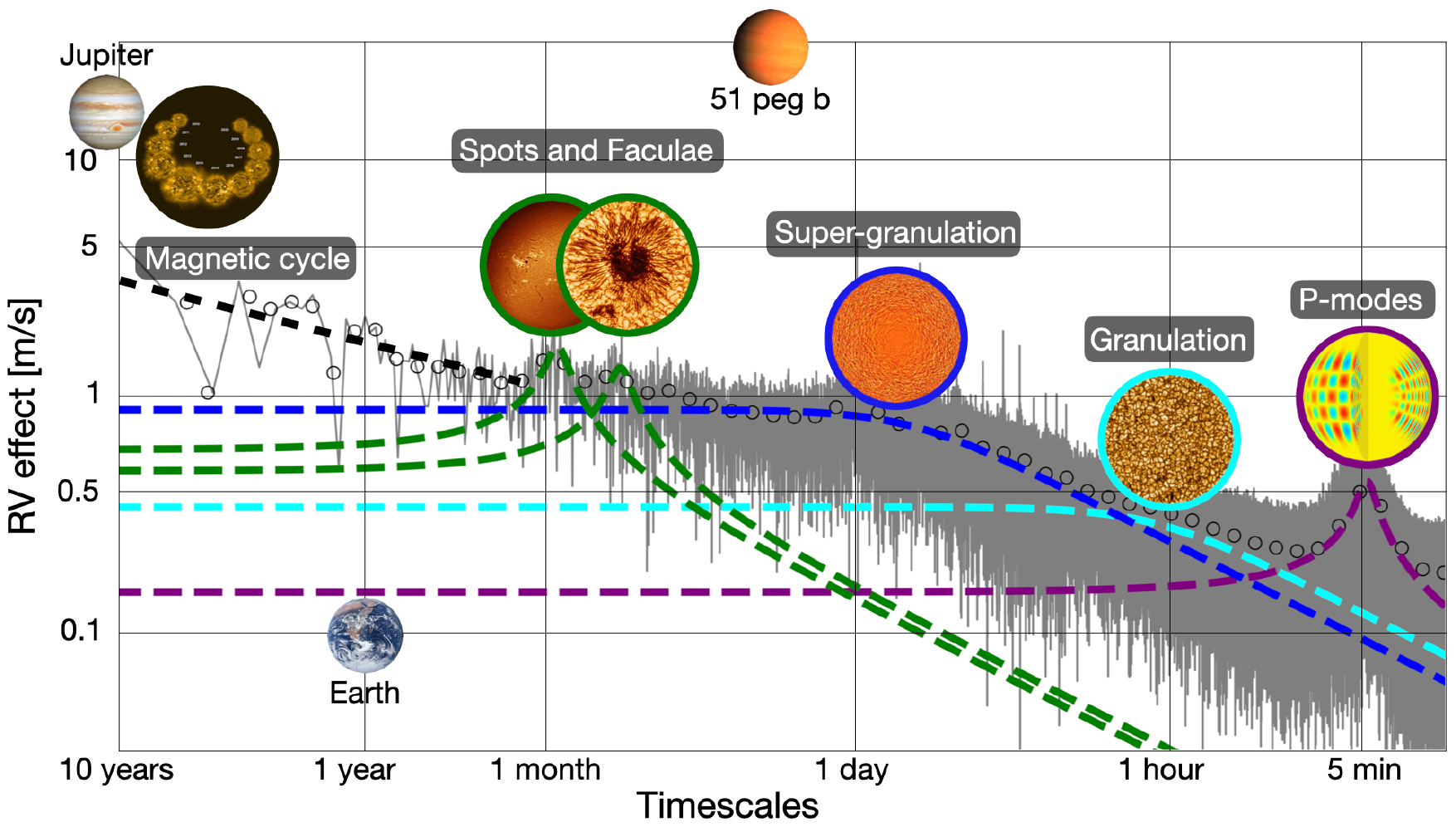}
\caption{Summary of the approximate timescales and RV semi-amplitudes of the stellar phenomena that most directly impact EPRV science efforts for a Sun-like star, with the RV semi-amplitudes of Earth, Jupiter, and 51 Peg b for reference. These phenomena span timescales from minutes to decades and RV semi-amplitudes of c\ms\ to dozens of \ms\ which poses significant challenges for designing RV survey and analysis techniques that can accurately sample, model, and mitigate their impacts. The background grayscale image represents the solar power spectrum density from \citet{AlMoulla:2023aa} with fitted contributions for the different stellar phenomena as colored dashed lines. Adapted from \citet{AlMoulla:2023aa} and a figure concept from Ryan Rubenzahl.}
\label{fig:stellar_variability}
\end{figure}

The RV technique is an indirect detection method whose goal is to observe a star and identify the tiny gravitational pull of an orbiting exoplanet. The measured RV is thus inherently contaminated by spurious stellar RV variations induced by surface flows and magnetic activity, which are summarized in Figure~\ref{fig:stellar_variability}.

\subsubsection{Surface flows} \label{surf_flows}

Many different types of flows, with different spatial scales and acting on different timescales, are at play in stellar photospheres. Stellar oscillations, granulation, supergranulation and meridional circulation all have an impact on RVs. The amplitudes of these flows are related to the star's gravity and effective temperature, and their RV contributions will be large for late-F dwarfs and decrease to non-significant perturbations for early-M dwarfs \citep[e.g.,][]{Dumusque-2011a,Guo:2022aa,AlMoulla:2025aa}.

Stellar oscillations in solar-type stars, also known as pressure modes (p-modes), are standing acoustic waves that drive a dilatation and contraction of the star's external envelopes on timescales of dozens of minutes for late-F dwarfs, to less than a minute for early M-dwarfs \citep[e.g.][]{Schrijver-2000, Broomhall-2009}. The RV jitter induced by the combination of all p-modes, ranges from a \ms\ for late F-dwarfs to ten c\ms\ for early M-dwarfs \citep[e.g.][]{Arentoft-2008, Chaplin:2019aa, AlMoulla:2023aa}.

\begin{marginnote}[]
\entry{jitter}{value that accounts for instrument systematics or stellar variability contributions that appear uncorrelated due to, e.g., the data set cadence}
\end{marginnote}

Granulation is a small-scale convective pattern, which manifests as 1000 km cells with a strong intensity contrast (15\%) and strong flows, at the level of 1 k\ms. The typical lifetime of granules on the Sun is 5 to 10 minutes, and averaging the velocity of those granules over the entire solar surface leaves a residual jitter at the level of 0.3 to 0.4 \ms on the Sun \cite[][]{AlMoulla:2023aa,Lakeland:2024aa} and even lower on K-dwarfs \citep[23 c\ms\ for HD166620][]{John:2025aa, Figueira:2025aa}. Magneto-hydro-dynamical (MHD) simulations converge to similar values \citep[e.g.][]{Cegla:2019aa, Sulis:2020aa}. 
Granules, where hot plasma is transported to the top of the photosphere, are surrounded by intergranular lanes, where cool plasma sinks back into the photosphere. Due to the temperature difference of the transported plasma, granules are much brighter than the intergranular lanes and thus dominate the RV signal. As the plasma in granules is moving toward the observer, from deep into the star to the top of the photosphere, the photosphere appears blue-shifted compared to the solar rest-frame, with a velocity of about 300 \ms\ for the Sun \citep[e.g.,][]{Dravins-1981}. This effect is known as convective blue-shift.

Supergranulation corresponds to a much larger convective pattern, with cells 30,000 km in diameter that evolve on timescales of up to 2 days for the Sun, and that present a slow vertical and a strong horizontal flow component \citep[20-30 \ms\ and 300–400 \ms\, respectively,][]{Rincon:2018aa}. Supergranules do not exhibit any contrast difference with the photosphere and are detected as a Doppler anomaly with respect to stellar rotation outside of the disc center, due to their strong horizontal flows. Like for granulation, the average over all supergranules leads to an RV jitter measured to be between 0.7 and 0.9 \ms\ on the Sun \citep[][]{AlMoulla:2023aa,Lakeland:2024aa}. Due to its non-trivial amplitude and challenging timescale, which cannot be averaged over within a night, supergranulation is a significant obstacle to the detection of Earth-analogues \citep[][]{Meunier:2020ab}.

The last important flow that is expected to affect RVs but that has not yet been measured is meridional flow \citep[][]{Makarov-2010}, a global-scale flow related to the conservation of angular momentum in the presence of differential rotation. The simulated impact on RV is dependent on the stellar inclination, with a larger effect for pole-on stars ($\sim$1.5 \ms) than for edge-on stars like the Sun \citep[$\sim$0.6 \ms][]{Meunier:2020aa}. Meridional flows are expected to vary on time scales similar to magnetic cycles and should decrease the long-term correlation between RV and magnetic activity index. However, this is not yet supported by a decade of HARPS-N solar observations (Dumusque et al. 2025, submitted).

\subsubsection{Magnetic activity} \label{magn_act}

Strong magnetic fields in the stellar photosphere are the origin of dark spots and bright faculae. As spots and faculae present a contrast difference with the photosphere, their presence on the solar surface will break the symmetry between the blue-shifted approaching and red-shifted receding hemispheres of a rotating star. Spots and faculae move across the visible surface of the star as it rotates, inducing a semi-periodic RV signal with a period linked to stellar rotation \citep[e.g.][]{Saar-1997b}. This RV effect induced by the contrast of active regions is often referred to as the flux effect. On the Sun, the contrast of spots is much larger than the contrast of faculae, however, faculae regions are much more extended than spots. This results in similar induced RV amplitudes of a few \ms. However, due to the opposite contrast of spots and faculae, their respective induced RV will be in opposition of phase, thus mitigating the total impact \citep{Meunier-2010a}.

\begin{textbox}[h]\textbf{Spots, Faculae, and Plage:}
Star spots are dark magnetic regions that can be up to a thousand degrees cooler than the surrounding photosphere because of their intense and large-scale magnetic fields which block convection, thereby reducing heat transfer. Faculae are bright magnetic regions caused by very locally concentrated magnetic fields ($\sim$100 km) that reduce the density of the surrounding plasma, making it more transparent and thus allowing an observer to see deeper and hotter photospheric layers. Both can cause significant variations in the RV signal of a star, as described above. In the RV literature the term plage is often used interchangeably with faculae to describe bright photospheric magnetic regions. But while the two types of regions originate from the same phenomenon, plage are actually located in the chromosphere.
\end{textbox}

Spot and faculae are also at the origin of another spurious RV signal at the stellar rotational timescale. The strong magnetic fields present in active regions suppress stellar convection, inducing a localized inhibition of the convective blue-shift (Sect.~\ref{surf_flows}). The region covered by a spot or a facula will therefore be red-shifted compared to the stellar photosphere, which will induce an RV effect. This convective blue-shift inhibition effect is impacted by spots and faculae in the same way, however, the effect from faculae will dominate as those regions are larger and much brighter than spots.  

Because the velocity of convection increases with physical depth into the photosphere, from 0 \ms\ at the top of the photosphere to 300 \ms\ deep inside for the Sun, spectral lines formed at different physical depths experience different inhibitions of convective blue-shift \citep[][]{Meunier:2017aa, Dumusque:2018aa, Cretignier:2020aa, Siegel:2022aa}. Shallow lines, at first order formed deep into the photosphere, undergo a strong inhibition of convective blue-shift due to magnetic activity. Very strong lines, formed across the entire photosphere, are less affected overall. 
By measuring the temperature of formation (a proxy for physical depth) of each wavelength bin in the spectrum and then measuring the RV of different shells in depth into the photosphere, it has been demonstrated that the RV effect induced by stellar activity varies significantly with line depth \citep[][]{AlMoulla:2022aa,AlMoulla:2024aa}.

When considering both the flux and convective blue-shift effect of active regions, the effect from the inhibition of convective blue-shift in faculae will dominate for the Sun, mainly due to the high velocity of convective flows, the larger size of faculae with respect to spots, and the slow solar rotation. On stars that rotate faster or that have slower convective flows such as M-dwarfs \citep[e.g.][]{Liebing:2021aa}, spots will start to dominate with their flux effect \citep[see Fig. 7 in][]{Dumusque-2014b}.

Finally, spots and faculae also induce an RV effect on the timescale of magnetic cycles. The Sun and solar-type stars are known to exhibit magnetic cycles with periods of several to a dozen years \citep[e.g.][]{Lovis-2011b}, characterized by a varying number of active regions on the stellar surface, from zero up to a few hundred in the case of the Sun. In addition to rotational modulation, these active regions constructively inhibit the star's convective blue-shift. As a result the star will become more red-shifted, inducing a strong correlation between the long-term RVs and the number of active regions for the Sun or the log(R'$_{HK}$) for other stars (Sect.~\ref{act_index}). The amplitude of the effect depends on the strength of the magnetic cycle, but also on the velocity of convection. Thus, for a magnetic cycle of similar amplitude, late-K dwarfs will show a smaller RV amplitude than early-G dwarfs \citep[10-20 \ms\ for the Sun, see equation 9 and 13 in][]{Lovis-2011b}. For M-dwarfs the RV signal generated by the magnetic cycle can be smaller still, with the cycles sometimes visible only in the stellar activity indicators or in photometry but not in the RVs themselves \citep{GonzalezHernandez2024,SuarezMascareno2025}.

\subsection{Mitigating Stellar Variability in the Data \label{sec:Mitigation}}

\subsubsection{Stellar oscillations}
By observing stars with an exposure time longer than the typical timescale of oscillation, which can be accurately predicted from scaling laws \citep[e.g.][]{Chaplin:2019aa}, it is possible to average out oscillations down to a few dozens of c\ms\,.

\subsubsection{Granulation and supergranulation}
Due to its relatively short timescale (5-10 minutes), observations can be designed with exposure times long enough to average out over granulation. Unfortunately, this strategy has proven inefficient \citep[][]{Dumusque-2011a,Meunier-2015} as when analyzing stellar and solar RVs \citep[][]{Dumusque-2011a,AlMoulla:2023aa}, the granulation signal appears at a timescale closer to one hour. This signal could potentially be induced by mesogranulation, a pattern predicted from simulations \citep[][]{Rast:2003aa, Bushby:2014aa}, but its detection on the Sun is still debated. Although uncertain in origin, this hourly signal is present in RV observations and taking several measurements per night separated by a few hours can help to mitigate it \citep[][]{Dumusque-2011a,Meunier-2015,Gupta:2024aa}. 

At longer timescales of up to two days, we are sensitive to supergranulation, with a measured jitter of 0.7 to 0.9 \ms. Averaging nightly measurements over several days helps slightly in mitigating this signal but \cite{Dumusque-2011a} and \cite{Meunier-2015} show that binning RV data over 10 days only reduces the impact by a factor of two. 

\subsubsection{Magnetic activity} \label{magn_activity}

Magnetic activity is \textit{the} major limitation to the detection of planetary signals due to its large variability in amplitude and it presenting on timescales similar to planetary orbits, from a few days for fast rotators up to a hundred of days for quiet M dwarfs. Magnetic activity also induces large RV signals correlated with stellar magnetic cycles, complicating the detection of planets in the habitable zones of sun-like stars. 
    
\textbf{Linear combination of activity proxies:} An efficient way to model the long-term effect induced by a magnetic cycle is to fit a linear combination of activity proxies. Generally, the Log(R'$_{HK}$), the FWHM, and the contrast of the CCF (or some combination thereof) are well correlated to the RV effect induced by magnetic cycles \citep[e.g.][]{Dumusque-2011c}. Generally, a smoothed version or one or more activity indexes are fitted linearly to the RVs \citep[e.g.][]{Dumusque-2011c,Udry:2019aa}. A recent study demonstrates that a linear correlation with the magnesium II activity proxy (0.99 correlation with Log(R'$_{HK}$)) allows us to decrease the long-term RV variation of the Sun due to the solar cycle from 2.95 to 0.41 \ms\,(Dumusque et al. 2025, submitted). On the rotational timescale, we can use the activity proxies cited above, in addition to the BIS SPAN of the CCF, to probe if any significant signal in the RVs could be explained by magnetic activity. This is usually done through a periodogram analysis. On that timescale, however, the interplay between the flux and convective blue-shift effects of the active regions weakens any linear correlations, which seems to depend strongly on the configuration of active regions on the solar surface.

\textbf{Use of Gaussian Processes:} The majority of RV studies where rotationally modulated magnetic activity is a nuisance use a Gaussian Processes (GP) framework to mitigate its impact (Sect. \ref{sec:GPs}). We note that GPs are very effective when a planet's period and phase are already strongly constrained by transit. In such a case, we can define strong priors for the planet parameters and leave the GP model what is left in the data. In cases without strong planet priors, modeling RV-only data with a GP runs the risk of significant overfitting. A recent development is the adoption of multi-dimensional GP models, based on the idea that the activity induced signal in RV can be modeled as a combination of the stellar flux (or any other activity proxy) and its gradient \citep{Aigrain2012}. This approach builds an underlying covariance that is then projected into the data as a linear combination of itself and its gradient \citep{Rajpaul2015}, tightening the relation between the RV GP model and the stellar activity proxies and reducing the risk of overfitting compared to traditional approaches. 

\textbf{Tailored line selection:} \cite{Thompson:2017aa}, followed by \cite{Dumusque:2018aa}, \cite{Wise:2018aa} and \cite{Cretignier:2020aa} all demonstrated that some spectral lines are more sensitive to magnetic activity than others. By measuring the RV of each individual line and then searching for either correlations with an activity proxy or periodic variations at the rotational timescale, it is possible to identify the more magnetically sensitive lines. Removing those lines and averaging the RV of the remaining, activity insensitive, lines allows us to significantly mitigate stellar activity by a factor of two to five \citep{Dumusque:2018aa, Lafarga:2023aa}. Spectral lines do seem to exhibit different levels of activity sensitivity from star to star, which requires doing a star-by-star analysis and thus require high S/N observations. We note that \cite{Larue:2025aa} seems to explain this star-by-star variation and might give clues on how to generalize this technique for any observations.

\textbf{Line-depth sensitivity:} After measuring the RV for different shells in depth in the photosphere, \cite{AlMoulla:2024aa} demonstrated that we can derive new magnetic activity proxies by taking the differential between pairs of these depth RVs. A linear combination of these new activity proxies, in addition to Log(R'$_{HK}$), allows a 20 to 30\% mitigation of magnetic activity. By developing a physically driven model similar to $FF'$ \citep[][]{Aigrain2012} using line-depth information, \cite{Siegel:2024aa} demonstrated a factor of two in magnetic activity mitigation by using their rotation convection framework.

\begin{marginnote}[]
\entry{spectral shell}{Compact representation corresponding to the spectrum normalized flux as a function of its gradient.}
\end{marginnote}

\textbf{PCA-based approaches:} Because it is challenging to fully understand how magnetic activity impacts spectral lines, pure data-driven approaches can help to find proxies that mitigate magnetic activity signals in RVs. In the SCALPELS framework \citep[][]{Collier-Cameron:2021aa}, an autocorrelation is first performed on a CCF time series to remove any contribution from a pure Doppler shift (i.e. a planetary signal). Then, a principal component analysis (PCA) is performed on the obtained products to find a basis for magnetic activity signals. Finally, the original CCFs are projected on this PCA basis and the time series of the obtained loadings (the coefficients fitted in front of each principal component) are used as activity indicators. In \cite{Collier-Cameron:2021aa}, the authors demonstrate that signal as small as 40 c\ms\ at 200 days can be recovered in three years of solar data. In \cite{Cretignier:2022aa}, the authors propose a new spectral representation called spectral shell which, opposite to a CCF, preserves the information regarding individual line depths, as this may be a critical parameter for modeling magnetic activity. A spectral shell can then be decomposed onto a Doppler-only shell and an activity shell. A PCA analysis of activity shells provides a basis where the loadings time series are used as activity indicators, like in the SCALPELS framework. The PCA spectral shell framework is also very efficient at mitigating magnetic activity signals on the rotational timescale \citep{Cretignier:2022aa}.

\textbf{Deep learning:} To probe for more complex non-linear relations, we can use deep neural networks. \cite{Beurs:2022aa} demonstrated that a convolutional neural network (CNN) can be used to model the effect of magnetic activity on the CCFs of HARPS-N solar observations. This model can be used to detect planets with amplitudes as small as 40 c\ms\ on a yearly period. \cite{Zhao:2024cc} developed a CNN framework that uses spectral shells as input and demonstrated an even better gain in magnetic activity mitigation, with detection limits as small as 20 c\ms\ for periods from a few to 500 days. Although this model works extremely well for the Sun, when applied to other stars the RV detection limit rises to K = 50 c\ms. This decreased detection threshold is likely due to the limited number of observations for stars other than the Sun, as a CNN trained on the Sun does not generalize well to other stars and a full retraining is necessary. \cite{Colwell:2025aa} tested the implementation of a CNN to line-by-line RVs and while only a few injection recovery tests were performed, this model was able to detect a K = 20 c\ms\ signal at a period of 50 days. 
The AESTRA framework \citep{Liang:2024aa} uses another approach, combining a convolutional Doppler shift estimator with an attentive autoencoder to probe magnetic activity at the spectral level. Although only tested on simulated data, AESTRA demonstrated extremely efficient recovery of planet signals in the presence of magnetic activity, detecting  amplitudes down to 10 c\ms\ on a 100-day orbit. 

\subsection{The Future} \label{future:stellar_var}

Although GP modeling at the RV level is very efficient at mitigating magnetic activity, it is unlikely that, even when using physically motivated kernels \citep[][]{Hara:2025}, the method will allow us to mitigate magnetic activity at the 10 cm/s level required to detect another Earth \citep[e.g.][]{Langellier:2020aa}. We therefore have to explore further how spectra are influenced by stellar variability and correct for it at this level.

Enormous progress has been made over the last decade in the physical understanding of how magnetic activity affects RV measurements through changes at the level of the spectrum. This progress led to the development of methods considering line depth as an important proxy for mitigating magnetic activity. But line depth is only the first order approximation and the community needs to continue investigating the effect of magnetic activity at the spectral level. Purely data-driven approaches on low-dimension spectral representations (CCF or spectral shells), like PCA or deep learning, currently provide the most efficient way of mitigating magnetic activity, however, they provide activity proxies that are not necessarily representative of the underlying physics. This cause deep learning models to have difficulties in generalizing to stars other than the Sun, and stellar data are generally not numerous enough to allow for further optimization of a pre-trained neural network using transfer learning. \cite{Zhao:2024aa} demonstrate a proof of concept on how to derive physically motivated activity proxies at the level of the spectrum and the community should investigate in similar directions. Those proxies could also come from analyzing disk-integrated solar observations obtained with HARPS-N \citep[e.g.][]{Dravins:2024aa}, NEID, EXPRESS, KPF or other EPRV instruments with solar feeds. PoET linked to ESPRESSO \citep[][]{Santos:2025aa} or the IAG solar telescope linked to a Fourier transform spectrograph will obtain high-resolution spectra covering the full visible band-pass for localized regions on the Sun containing spots, faculae or the quiet photosphere. Those data will be useful for identifying new stellar variability proxies at the spectral level which can then be used in a framework like YARARA to mitigate stellar activity at the spectral level \citep[][]{Cretignier:2021aa}. If properly understood, we could also develop specific kernels to directly model stellar variability at the spectral level using GPs \citep[e.g.][]{Yu2024}.

On the side of granulation, recent MHD simulations demonstrate that granulation has an impact on the bisector of spectral lines \citep[][]{Palumbo:2024aa}. Although not detectable on a line-by-line basis because the effect is too small compared to the limited RV precision on a single line induced by photon noise, modelization can help in knowing which lines to group together to increase RV precision and detectability. \cite{Palumbo:2024aa} predict that we could mitigate 30\% of the granulation signal using current instrumental resolution ($R\sim$100,000). We could potentially obtain better mitigation by going to even higher resolution ($R\sim$300,000), but that is challenging from an instrumentation point of view.

Supergranulation is the less studied and understood stellar variability component. On the side of observations, supergranulation can be probed in RVs \citep[][]{Dumusque-2011a, AlMoulla:2023aa, OSullivan2025} but not at the spectral line level due to tiny local effects. Contrary to granulation, MHD simulations that only work at the level of small atmospheric boxes cannot model the large-scale flows at the origin of supergranulation and thus simulations cannot guide us on which lines to combine to boost the signal. More information could come from combining data collected from solar telescopes connected to EPRV spectrographs all around the world.

Although not yet used in practice, the current best strategy to search for planets in the presence of granulation and supergranulation is to model these signals using a GP framework with appropriate kernels \citep[][]{OSullivan:2024aa}. However, we note that an analysis of 8 years of solar data demonstrated that the timescale of supergranulation varies by an order of magnitude with the solar magnetic cycle, complicating standard GP modeling efforts.

\section{MODELING PLANETARY SYSTEMS}

\subsection{RV Survey Considerations}
Designing a survey to detect the RV signatures of planets and produce robust Keplerian orbital solutions requires early consideration of the target star(s), the type of planet(s) the survey aims to detect, the capabilities of the RV instrument, and the scheduling flexibility of the facility that will be used. Observers should also consider the likely timescales and amplitudes of both the planets they seek to detect and the various types of RV variability expected from the host star. Calculation of the expected RV semi-amplitude of a planet is straightforward, aside from knowing the eccentricity of the planet, using Equation \ref{eqn:semiamp}. Accurate predictions of the RV signal due to stellar variability is more challenging but predictive relations based on the star's log(R'$_{HK}$) values or time series photometry have been put forth \citep{Santos2000,Aigrain2012,SuarezMascareno2018}. 

When following up transiting planet candidates with well defined orbital periods that \textit{a priori} knowledge can be used to design RV campaigns that efficiently sample the candidate's orbital phase curve \citep{Burt2018, Lam2024}. In uninformed surveys with no specific planet candidates but rather the goal of discovering planets in a given mass/period parameter space, the surveys can instead be optimized to reach a certain RV semi-amplitude sensitivity as a function of orbital period \citep{Gupta:2024aa}. Examples of large scale, modern EPRV surveys include the HARPS-N Rocky Planet Search \citep{Motalebi2015}, The EXPRES 100 Earths Survey \citep{Brewer2020}, the NEID Earth Twin Survey \citep[NETS,][]{Gupta2021, Gupta2025}, the HUnting for M Dwarf Rocky planets Using MAROON-X \citep[HUMDRUM,][]{Brady2024} survey, the ESPRESSO GTO Survey \citep{Pepe:2021aa}, and the HARPS3 Terra Hunting Experiment \citep[THE,][]{Hall2018}.

\subsection{Deciding Upon a System Model} \label{stellar_mod}

Identifying potential exoplanet signals in an RV time series often starts with the creation of a periodogram, a Fourier-like power spectrum that highlights potential periodic signals within a time series. Generally speaking a periodogram compares the log-likelihoods of a base model such as Gaussian white noise against a model that includes both the white noise and a periodic signal of frequency $\omega$ across a user-specified frequency grid. Significant peaks in a periodogram may correspond to planets with those orbital periods and can help astronomers decide which candidate signals are worth examining in more detail. 

\begin{marginnote}[]
\entry{False Alarm Probability}{quantifies the probability that random noise alone could produce a peak in the periodogram as high as the  observed peak.}
\end{marginnote}

RV planet searches traditionally use a Lomb-Scargle periodogram \citep{Lomb1976, Scargle1982, VanderPlas2018} wherein the periodic signal is a sine function representing the RV variations of a planet on a circular orbit. If a peak in the periodogram looks promising one must then decide how likely it is that the peak is due to an astrophysical signal such as an exoplanet's gravitational tug. A traditional metric for this is the False Alarm Probability \citep[FAP,][]{Baluev2022}. The FAP asks the question ``if there is no planet present, then how unlikely is this peak given the data?'' and a general rule of thumb is that FAP values below 0.1\% are worthy of additional analysis and those below 0.01\% indicate a true periodicity (albeit not necessarily one due to an exoplanet!) in the data set. Improvements to the original Lomb-Scargle approach have included fitting a Keplerian model rather than a sinusoid \citep{ZechmeisterKurster2009}, simultaneous consideration of multiple signals \citep{Baluev2013}, and Bayesian formalisms utilizing marginalized likelihoods \citep{Mortier2015}. 

\begin{marginnote}[]
\entry{False Inclusion Probability}{considers models with different numbers of planets and calculates the posterior probability of the presence of a planet using nested sampling.}
\end{marginnote}

A more modern and Bayesian approach is the False Inclusion Probability \citep[FIP,][]{Hara2022} periodogram which uses the data and a set of user-assigned priors to instead ask ``How probable is it that a planet orbits at this period given the RV measurements and my additional knowledge of the system?''. While the FIP is better equipped to handle aliasing, the presence of multiple planets, and noise model uncertainties, it is also influenced by the priors assigned by the user regarding the number of planets, the type(s) of noise present, etc, and so additional care must be taken in these decisions. 

\begin{textbox}[h]\textbf{Concerning Aliases:}
Data from an infinite, noise-free, RV time series would produce a delta-function at the orbital frequency of the planet in Fourier space. Real astronomical observations, however, are finite, noisy, and unevenly sampled, which leads to more complex periodograms. The impact of recurring or irregular gaps in the data can be captured by computing the Fourier transform of a binary time series with values at the timestamps of observations set to 1, known as the window function \citep{Roberts1987}. If the RV data set contains a peak caused by a planet with a frequency of $f = f_{p}$ then a peak in the window function with a frequency of $f = f_{w}$ will produce aliases of the planet's signal at $f = n|f_{p}\ \pm\ f_{w}|$ in the periodogram. As aliases can add either constructively or destructively, one cannot assume that the highest peak in a periodogram is necessarily the true period of the planet. See \citealt{Dawson2010} for an overview of aliases in RV data and how to distinguish a planet's true orbital frequency from its aliases. 
\end{textbox}

Once one or more candidate signals are identified, it is time to develop a full RV model for the system. Here we will assume that some aspect of the host stars variability is still present in the data, though efforts to remove any correlated signals during post processing are becoming increasingly common as discussed in Section \ref{sec:Mitigation}. A number of publicly available software packages exist for this purpose (see Section B of the Appendix), many of which address not only the planet's RV parameters but also some form(s) of stellar variability, related transit photometry data, and even the stellar parameters of the host star. 

\textbf{Overview of RV model components:} The orbital elements used to describe a planet's Keplerian signature when modeling RV data are: the orbital period ($P$), eccentricity ($e$), RV semi-amplitude ($K$), longitude of periastron of the star's orbit ($\omega$) and either the time of periastron (T$_{p}$) or the time of inferior conjunction (T$_{c}$). In addition to the Keplerian orbital elements, the RV model can also contain terms for a linear ($\dot{\gamma}$) or quadratic ($\ddot{\gamma}$) acceleration. These terms can capture if the star is moving along an orbit with a period much longer than the observational baseline such the portion of the orbit that is discernible from the data does not look like a Keplerian but rather like a positive or negative trend (the linear acceleration) if the star has only been captured between the two RV quadrature phases, or like a parabola (the quadratic acceleration) if the star has passed through one of the RV quadrature points but not the other. If included in the model and found to be statistically significant, these trends can serve as an early indication that the star has a long-period stellar or planetary companion.

The model should also contain an RV offset ($\gamma$) and RV jitter ($\sigma$) term for each instrument included in the fit, and sometimes for each distinct era of the same instrument if its behavior or performance changed due to hardware upgrades or other physical interventions. The measurements provided by modern RV spectrographs are relative velocities, and so they do not share an absolute zero point. Fitting for an instrument-specific offset removes the impact of this lack of shared reference frame, but the offset can be challenging to constrain if the data sets do not have any temporal overlap. The jitter term captures the effect of noise sources that are not well resolved in the model and which increase the scatter in the data. These can be contributions from the instrument itself (e.g., instrument drifts due to thermal cycles) or the star (e.g., stellar granulation and granulation happening over a timescale of multiple hours that cannot be effectively modeled and removed from data with a nightly cadence). By adding the the jitter in quadrature to the RV errors, like another white noise source, we normalize the model likelihood and make it consistent with the observed scatter in the data.

\textbf{Selecting a fitting basis:} The Keplerian model parameters can be analytically transformed into a variety of orbital parameter basis sets with different characteristics that help to increase the rate of convergence or decrease biases. For example, we know that the posterior distribution of the eccentricity is not well sampled for orbits with small eccentricities \citep{LucySweeney1971}. We can therefore adopt a parameterization that fits for $\sqrt e \sin(\omega)$ and $\sqrt e \cos(\omega)$, rather than just $e$ and $\omega$, as this prevents truncation when the eccentricity is close to zero. See \citealt{Fulton2018} for a useful overview of different orbital parameterizations used in RV modeling and \citealt{Eastman2013} for a detailed discussion of the implicit priors imposed on the orbital eccentricity and longitude of periastron by the choice of fitting basis. There is no single default basis for modeling a planet's Keplerian signal but the P, T$_{p}$, $\sqrt e \sin(\omega)$, $\sqrt e \cos(\omega)$, K basis is often used as a starting point as it produces flat priors on each orbital element and helps to speed MCMC convergence \citep{Fulton2018}. When finalizing the RV model for a given data set it is a good rule of thumb to fit the data using several different basis sets and check that this does not produce statistically significant changes in the posterior distributions of the orbital elements. 

\textbf{Assigning Priors:} Most fitting packages allow the user to assign informative priors or uninformative boundary ranges to the orbital and instrument elements in the model. When setting these values, the user should try to to balance physics-based constraints with model flexibility so that the model does not artificially drive the results. Priors on orbital parameters like period and orbital phase should reflect what is known from auxiliary data (e.g., if the planet is seen to transit then its period is likely known to high precision) but not be so restrictive that they exclude plausible aliases. Note, however, that if the model is a joint fit that includes both transit and RV data, you should not place priors on the period and/or orbital phase originally derived from the transit data as this will double count the transit information and produce unjustifiably small uncertainties \citep{Eastman2019}. The RV semi-amplitude and orbital eccentricity priors can be set to avoid non-physical solutions with negative semi-amplitudes or eccentricities greater than one, but users should consider the choice of prior distribution to avoid biasing these values towards artificially high values \citep{Eastman2013, Stevenson2025}. For stellar variability-focused terms such as stellar jitter or Gaussian Process hyperparameters (discussed below), priors should be wide enough to encompass the expected astrophysical variability based on independent data such as photometric time series or log(R'$_{HK}$) measurements, but not so broad that they push the sampler into uninformative regions. In general, users should always test their model's sensitivity to the choice of priors before deciding upon the final result to ensure that the inferred planet parameters are data-driven rather than prior-driven.

\begin{marginnote}[]
\entry{Bayesian evidence}{probability of the data given a particular model after accounting for all possible parameter values within that model}
\end{marginnote}

\textbf{Deciding how many planets belong in the system model} : The most principled approach to select between models with different numbers of planets is a Bayesian model comparison, which involves computing the marginal likelihood (the \lq{}Bayesian evidence\rq{}) for an N-planet model versus an (N+1)-planet model. The ratio of the Bayesian evidence for two different models is known as the Bayes factor \citep{KassRaftery1995} and quantifies the statistical evidence in favor of one model over another. Accurate determination of the Bayes evidence can be challenging and computationally demanding \citep{Nelson2020} and so many RV fitting packages instead make use of the Bayesian Information Criterion, a computationally simpler approximation for the log of the Bayesian evidence that works best with large sample sizes. Models with lower BIC values are preferred and when deciding whether to include an additional planet in the model for an RV data set the community generally requires a $\Delta$BIC value of at least 5 to adopt the N+1 planet model. The BIC is a conservative metric, so as more data is added it becomes harder to justify the inclusion of additional planets. But as the data set grows, the BIC will tend to select the true model if it is included in the list of models being compared. 

Users might also consider the Akaike Information Criteria (AIC) as a model selection metric. The AIC is more permissive and will more readily include additional planets which can lead to overfitting in sparse data sets \citep{Chakrabarti2011}. It does not, however, assume that the true model of the system is present in the list of comparisons and instead selects the model that best predicts future observations of the system. As we never know the full extent of a star’s planetary system nor its stellar variability when developing RV models, this can serve as a more realistic representation of our knowledge of the system.

A large $\Delta$BIC or $\Delta$AIC value alone does not guarantee the presence of an additional planet in the system, especially in cases where the physical or statistical models used are inaccurate \citep{HaraFord2023}. And even beyond its statistical robustness there are a number of additional criteria that the candidate signal should meet before it is put forth as a confirmed planet in the literature. These include, but are not limited to:
\begin{itemize}
    \item Is the signal sufficiently uncorrelated with simultaneous stellar activity metrics to suggest that it is not due to the star's variability?
    \item Is the signal unlikely to be driven by aliases induced by peaks in the window function?
    \item Are the signal's RV semi-amplitude and orbital period steady over many orbits?
    \item Is the signal robust to different RV fitting basis and model prior selections?
    \item Is the signal robust to the removal of small subsets of the RV data?
\end{itemize}

\subsection{Use of Gaussian Processes in RV Models} \label{sec:GPs}

Gaussian Processes (GPs) are a type of statistical model that provide a powerful yet flexible framework for modeling the complex, time-correlated, signatures of stellar phenomena in RV data. At present, GPs are most often used to model the impacts of rotational modulation in RV time series \citep[see, e.g.,][]{Barragan2022b, Klein2024}, though recent works have also applied them to super-granulation \citep{OSullivan2025} and magnetic cycles \citep{Basant2025}. Gaussian processes are defined by a mean and a covariance function and while we will not endeavor to describe the nuances of GP Regression here (see instead \citealt{Aigrain2023}) a few key components to keep in mind are:

\begin{marginnote}[]
\entry{kernel / covariance function}{parametric function that quantifies the correlation between any two data points, controlled by a set of parameters known as hyperparameters}
\end{marginnote}

\begin{enumerate}
    \item When modeling RV data, the mean function encodes information about the Keplerian signals imparted by any planets included in the model
    \item The covariance function (often referred to as the GP's kernel) sets the covariance between pairs of data points and captures the temporal correlations and characteristic timescales of the star's variability \citep{Roberts2013}.
    \item Each GP kernel has a set of corresponding hyperparameters that describe its characteristic behavior and the properties of the correlated stellar variability. These hyperparameters can sometimes be tied to physical properties of the star and be informed by complementary observations \citep{Nicholson2022}. 
\end{enumerate}

\begin{figure}[h]
\includegraphics[width=5in]{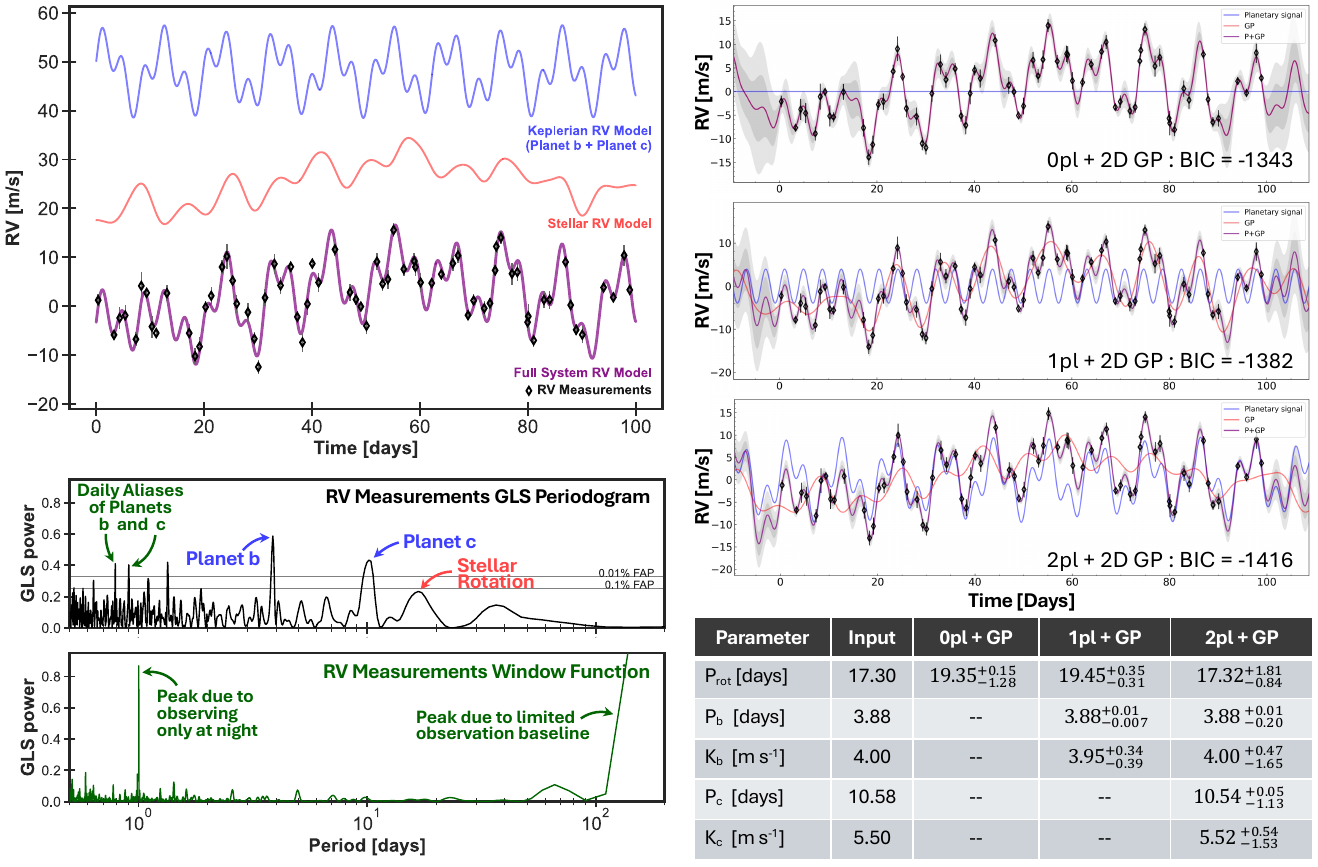}
\caption{Key steps in the exoplanet discovery process. Top left: RV model of a star (purple curve) that includes two planets (blue curve) and rotational modulation from spots (red curve). Black diamonds represent the measured RVs and associated uncertainties of a simulated observing campaign that incorporates realistic target availability from a single observatory. Bottom left: generalized Lomb-Scargle periodogram and window function of the simulated RV measurements with labels for key features corresponding to the planets, their daily aliases, and the stellar rotation. Right panel: Results for three fits to the data each including a two dimensional quasi-periodic GP based on the RV measurements and associated log(R'$_{\rm{HK}}$) measurements and then a 0-, 1-, or 2-planet model. While the GP can increase its complexity to absorb the signatures of one or both planets, both the incorrect stellar rotation periods and the larger Bayesian Information Criteria of the 0- and 1-planet models make clear that the 2-planet model is preferred. All data and fits were generated using Pyaneti \citep{Barragan2022a}.}
\label{fig:planetfitting}
\end{figure}

\begin{marginnote}[]
\entry{hyperparameters}{kernel-specific values that quantify and describe the properties of the correlated stellar variability}
\end{marginnote}

The quasiperiodic kernel is often used to model rotational modulation as its functional form allows for the evolution of spot patterns on timescales longer than the stellar rotation period. In cases where the star also exhibits power at harmonics of the rotation period, the use of either a quasiperiodic + cosine kernel or a combinations of stochastic harmonic oscillators with periods set to $P_{rot}$, $P_{rot}$/2, etc, have proven successful \citep{Perger2021, SuarezMascareno2025}. The similarity in the shapes and timescales of planet- and activity-induced RV variations results in a risk of the GP \lq{}absorbing\rq{} some portion of an RV planet signature and underestimating the planet's signal. Indeed, some studies have noted that the choice of whether to include GPs in the system model and the choice of which GP framework to adopt can produce significant variations in the best-fit orbital eccentricity and RV semi-amplitude \citep{Osborne2025, Tran2024}.

To help avoid such model-driven variance, users can model the simultaneous shared behavior between RV measurements (which are influenced by both the planet(s) and the star) and stellar activity metrics (which are influenced only by the star) using a multi-dimensional GP framework that relates activity-induced RV variations to activity sensitive metrics through the use of a latent variable based on the star's spot coverage fraction \citealt{Rajpaul2015}. This framework allows the stellar activity component of an RV time series to be constrained and disentangled from the Keplerian components. Even when carrying out multi-dimensional fits that place constraints on the stellar variability it is still valuable to perform some level of cross-validation testing to ensure that the preferred GP model is not overfitting the data \citep{Blunt2023}. 

While GP models generally improve detection robustness they also add non-trivial computational costs to any exoplanet fitting effort; the cost of evaluating the likelihood of a GP is proportional to $n_{obs}^{3}$. When searching for planets with smaller RV signatures, which require larger data sets, these scalings can make the modeling of stellar variability intractable. In response, semiseparable and/or sparse noise models have been developed to capture major aspects of rotational modulation and calibration while reducing the computational overhead to only a linear dependence on $n_{obs}$ \citep{Foreman-Mackey2017, Delisle2022}.

\subsection{The Future} \label{future:sys_model}

Accurate modeling of exoplanet systems begins with intentional survey design. EPRV surveys need to sample the time scales relevant to both the planets they seek and the stellar magnetic phenomena they need to mitigate \citep{Gupta:2024aa}. Sustained, high cadence observing will likely require dedicated facilities and/or coordinated efforts between observatories around the globe. Suggestions of what RV sensitivity can be achieved via dedicated observing effort can be seen in instrument-specific solar RV analyses \citep[e.g.,][]{Klein2024,Ford2024} and multi-instrument solar RV comparisons \citep{Zhao2023}. 

Coordinated, multi-instrument observing efforts will allow for the behavior of the spectrographs themselves to be included as parameters in the global model. Signals that are shared between data sets can be attributed to the common target star while signals appearing in only one instrument's data are tied to that particular facility and/or pipeline. This can mitigate the risk of, e.g., slow temperature drifts in certain optical elements producing changes in the instrument profile that can masquerade as long-period stellar variability \citep{SuarezMascareno2023}. Care should be taken to ensure that multi-instrument fits treat the models for different instruments as correlated (while allowing each instrument to have its own GP amplitude, jitter, and $\gamma$ offset term) rather than only requiring that the instruments share GP hyperparameters while treating the data sets themselves as uncorrelated \citep{Blunt2023}. The EPRV community is finalizing a new data standard (\url{eprv-data-standard.readthedocs.io}) to lower the barrier to such combined analyses.

Physically motivated kernels that represent stellar activity as a stochastic process \citep{Hara:2025} have demonstrated better predictive accuracy than traditional GPs and better cross-validation results, reducing the risk of overfitting. And efforts to model the time series of the CCFs, rather than just the RVs, show promise for more robust identification between activity and planetary signals \cite{Yu2024}. Continued investigations into how best to model both shorter (e.g., super-granulation) and longer (e.g., magnetic cycles) period stellar phenomena will be an important focus as more EPRV surveys begin releasing multi-year data sets designed to detect temperate, rocky planets \citep[e.g.,][]{Meunier:2020ab}. In particular, developing models that track how these different phenomena influence one another -- such as how the number of star spots fluctuates as the star moves through its magnetic cycle -- will be necessary when developing models for 5-10 year EPRV time series.

\section{FUTURE ADVANCEMENTS IN EPRV SCIENCE}

State-of-the-art EPRV spectrographs deliver 20-30 c\ms\ photo-noise precision on single measurement and 10 cm/s stability on a nightly basis. On longer timescales, ESPRESSO has demonstrated that the RV jitter of the very quiet star $\tau$\,Ceti reaches $\sim$50 c\ms\ over five years of observation after magnetic activity mitigation \citep[][]{Figueira:2025aa}, providing an upper limit to long-term calibration stability. Such precision and stability open the door to the discovery of temperate, super-Earth planets around nearby, Sun-like stars. The field aspires towards even more challenging goals, however, such as the characterization of the true Earth analogs (K = 9 c\ms\ over 1 year periods) sought by PLATO and the Habitable Worlds Observatory, and the direct measurement of expansion in the Universe via the detection of red-shift drifts in distant galaxies (5 c\ms over a decade). Advancing our current capabilities to the levels required for such science feats will require a significant community effort that includes improved instrumentation, dedicated and intentionally designed surveys, and advanced data analysis techniques for both spectral extraction and spectral post-processing to mitigate stellar variability \citep{Crass2021}. Here we highlight some areas of particular importance for reaching $\leq$10 c\ms\ precision and stability: \\

\noindent \textbf{Instrumentation}: While significant progress has been made over the past decade, more effort is needed to push RV instrumentation to the level required to support c{\ms} stellar Doppler measurements. Achieving this goal will require developing more reliable, spectrally-broad frequency standards that provide both accurate and precise calibration. Many promising LFC technologies are being explored, and new FP cavity designs could help to bridge the gap. It will also be crucial to leverage existing RV facilities to gain a deeper understanding of individual instrumental systematics. This exercise will inevitably both influence the design of future systems and help prioritize future subsystem development activities. Exploring architectures for higher resolving power may be advantageous (or necessary) for stellar activity mitigation. This will likely require exploring new spectrometer design families, both seeing-limited (likely with new slicing approaches) and diffraction-limited. All of these technological advancements will be crucial to investigate as the community explores a new generation of replicable designs to develop a fleet of RV facilities that complement future space-based Earth analog searches \citep{Crass2021}. \\

\noindent \textbf{Data Reduction and Post Processing Pipelines}: Although EPRV instruments are extremely stabilized, any significant instrumental interventions, such as thermal cycles or spectrograph reopening, drastically modify the PSF and impact long-term RV stability. In addition, more subtle but still observable mechanical stresses and optical aging also modify the PSF slightly over years. The current solution to deal with such PSF variations is to recalibrate EPRV spectrographs on a daily basis. Even a perfect calibration will induce an offset in the RVs derived from a stellar spectrum, however, because changes in the PSF impact stellar spectra differently than they do wavelength solution calibration spectra \citep[][]{LoCurto:2015aa,Schmidt2021}. The solution is to actively model the PSF and de-convolve it (or forward model the spectrum) to obtain calibrations and stellar spectra devoid of these instrumental systematics. This can be done at the level of the extracted spectra, as explored in \cite{Schmidt2024}, or at the level of the raw frames, as proposed by the spectro-perfectionism framework \citep[][]{Bolton:2010aa}. Even ideal spectral extraction cannot model and mitigate all instrumental signals. Time series of extracted spectra, both from calibrations and stellar observations, convey substantial information regarding instrumental signal residuals. Correcting for the observed systematics at the level of extracted spectra, before measuring a wavelength solution or the RV of a stellar spectrum, can significantly improve calibration and RV precision \citep[see, for example, the YARARA framework][]{Cretignier:2021aa, Cretignier:2023ab}.\\

\noindent \textbf{Stellar Variability Mitigation}: Enormous progress has been made over the last decade to increase our understanding of how stellar variability affects stellar spectra. Even so, our knowledge is still very limited and more work is needed to obtain a deeper physical understanding. The community should continue to pursue in depth explorations of the disc-integrated solar data obtained by the HARPS-N, NEID, EXPRESS and KPF solar telescopes, and the resolved solar observations that PoET and the IAG solar telescope will deliver. This will allow us to find stellar variability proxies at the level of the spectra and then directly model stellar variability at the spectral level using a framework such as YARARA \citep[][]{Cretignier:2021aa}. Alongside these observational efforts, the field should leverage MHD simulations to increase our physical understanding of stellar variability. These simulations could help us to identify novel stellar variability signatures in real observations and to develop mitigation techniques that can be applied to any star. \\

\noindent \textbf{Surveys \& System Models}: Developing more accurate representations of exoplanet systems will require carefully designed surveys that produce data with the cadence and precision required to resolve the time scales and RV amplitudes of the stellar phenomena and exoplanets we aim to characterize. Coordinated, multi-instrument observing efforts can help to sample timescales that are challenging for a single facility and provide insights on whether specific signals in the data are due to the star or to systematics from an individual instrument \citep{Zhao2023}. Regardless of how many RV instruments are used, knowledge of their behavior should be considered as an additional set of parameters in the global model \citep{Blunt2023}. Gaussian Processes have proven to be a powerful tool for characterizing stellar variability, and the RV community is continually identifying new ways to apply them to planet fitting efforts. A continued focus on incorporating knowledge of stellar physics and the specific target star into the models \citep{Hara:2025} should help to reduce the risk of overfitting. Better model selection methodologies need to be developed for multi-dimensional GPs as the commonly used Bayesian and Akaike Information Criteria scale with $n_{obs}$ and are thus not comparable across models of different dimensions (RV vs. RV + FWHM, etc). Looking forward, development of kernels that consider not only multiple kinds of stellar variability but also any covariances between the different phenomena (e.g. between the phase of the star's magnetic cycle and the amplitude of its spot modulation) will be needed to more accurately capture the star's behavior over decade long timescales.

\section*{DISCLOSURE STATEMENT}
The authors are not aware of any affiliations, memberships, funding, or financial holdings that might be perceived as affecting the objectivity of this review. 

\section*{ACKNOWLEDGMENTS}
Part of the research was carried out at the Jet Propulsion Laboratory, California Institute of Technology, under contract with the National Aeronautics and Space Administration (NASA). X.D acknowledges the support from the European Research Council (ERC) under the European Union’s Horizon 2020 research and innovation program (grant agreement SCORE No 851555) and from the Swiss National Science Foundation under the grant SPECTRE (No 200021\_215200). This research has made use of the NASA Exoplanet Archive, which is operated by the California Institute of Technology, under contract with the National Aeronautics and Space Administration under the Exoplanet Exploration Program.

\newblock

\noindent The authors wish to thank Khaled Al Moulla, Ashley Baker, Oscar Barrag{\'a}n, Alison Duck, Steve Gibson, Tobias Schmidt, Alejandro Suárez Mascareño, Ryan Terrien, and Christian Schwab for their helpful comments and suggestions.

\newpage

\appendix{}

\begin{summary}[]
\section*{SUPPLEMENTAL APPENDIX}
\end{summary}

\section{Standard spectrograph calibrations}

Before extracting the flux from different traces while correcting for detector systematics, a series of specific calibration frames are required. Here we provide additional detail on the types of calibration frames that are utilized in modern RV data reduction pipelines. 

\newblock

\begin{marginnote}[]
\entry{master calibrations frame}{calibration image generated by stacking several similar frames to increase S/N, generally at least five frames. While stacking, an outlier-rejection algorithm (e.g., sigma-clipping) is used to reject cosmic ray hits or bad frames.}
\end{marginnote}

\textbf{Bias:} CCD images with no integration time, called bias calibration frames, are used to measure the signal induced by CCD electronics when reading the detector, as no light is reaching the array.
This signal is composed of a mean flux level, called bias, some large-scale residuals around this mean, called bias residuals, and some pixel-to-pixel statistical noise called read-out-noise (RON). Each bias frame will be affected by a slightly different non-zero mean flux level, called the bias, which is an electronically induced offset added to the CCD reading electronics to ensure that all pixel values are positive when converted to a digital number. This mean flux level is generally measured from specific overscan regions that are reads of the detector without shifting charges into the readout electronics (empty reads, see Figure 8). Once the mean bias is measured and subtracted, the corrected bias frame will still show some residual variations that evolve slowly with time. By building a master bias residual frame, to increase S/N and reject the systematics induced by cosmics, we can correct any frame for those residuals. The RON is generally measured by taking a rms of the difference between two consecutive bias frames (to remove the bias and bias residuals). Because acquiring bias frames takes very little time, master bias residual calibrations should be obtained every day. We note that all the calibration and science frames described below should be pre-processed by correcting for the mean overscan bias and the master bias residual.

\textbf{Dark:} A CCD image taken with a closed shutter and a long exposure time, called a dark calibration image, highlights thermal effects in the CCD, called the dark current. Although most pixels will present a similar dark current, some hot pixels will have much higher levels. Modern CCDs have very low dark current levels of only a few electrons per pixel per hour, and so each individual dark calibration frame must be taken with a long exposure time, generally an hour. Hot pixels evolve slowly with time, so new master dark calibrations are generally taken every month.

\textbf{2D flat field:} High-resolution EPRV spectrographs are generally equipped with LEDs or broadband lamps that uniformly illuminate the detector. 2D flat-field frames obtained with different exposure times are used to measure the pixel response and determine the gain of the detector. The gain, i.e. the inverse of the photoelectron to ADU conversion factor, is measured as the inverse slope of the photon transfer curve which corresponds to the flux variance as a function of the measured flux \citep[e.g.][]{Janesick:1985aa, Downing:2006aa, Astier:2019aa}. 
As the variance will follow Poisson statistics, it should be equal to the flux and therefore the slope should be unity. However, a photoelectron will not necessarily correspond to a value of one in Analogue-to-Digital units (ADUs), which is the unit measured by CCD electronics. The measured slope can therefore be used to convert ADUs to photoelectrons and therefore obtain flux in physical unit. 
Several master frames with different exposure times ($\ge3$) are used to measure the gain and detect bad pixels that present a non-linear response. We note that hot pixels detected with a dark calibration should be excluded when measuring linearity. Like for dark calibrations, bad pixels evolve slowly and so 2D flat-field calibrations are only required every month.

\textbf{Flat field:} Before extracting the flux from each trace in each spectral order (see Fig.~\ref{fig:spectral_format}) and correct for pixel-to-pixel variations and fringing effects, we first must localize those traces on the CCD. This can be done using frames in which a single trace per spectral order (or the different traces of one sliced fiber) is illuminated by a lamp providing a continuum spectrum, generally from a tungsten-halogen lamp, a laser-driven light source (LDLS) or a supercontinum source. We note that illuminating several traces at the same time is still possible but makes any localization algorithm more complicated. After the localization of the traces is done, a master frame built from such frames can be used to measure pixel-to-pixel variations and fringing effects, but also to measure the trace profile in cross-dispersion for each spectral order.
We note that here, at least 10 frames are required to build the master flat-field. This ensures that the noise when measuring pixel-to-pixel variations in the master flat-field calibration is $\sim 3$ ($\sqrt{10}$) times smaller than the noise in a science observation (considering similar signal-to-noise (S/N) between calibrations and science observation). In the case of solar observations, where dozens to hundreds of observations are taken everyday, the S/N of daily-binned spectra will be limited by the master flat-field S/N. Master flat-field calibrations need to be obtained daily. 
These flat-field frames are generally time-consuming as they need to be taken daily, and it is therefore tempting to illuminate all traces at once. However, we note that cross-talk between fibers (which can be measured if only one trace is illuminated at a time and amounts to up to 0.5\% in the case of ESPRESSO), contamination between traces and spectral ghosts will all degrade the quality of the master flat field and reduce the precision at which spectra are extracted, especially at low S/N.

\textbf{Stray light correction:} For all frames in which different traces will be illuminated, such as the flat-field calibrations discussed above and all calibration and science frames that will be discussed below, optical scattering inside the spectrograph will create a diffuse background halo on the raw frame. This light will contaminate the flux on the different traces and should be corrected for, mainly when observing at very low S/N. Generally, this contamination is first modeled by measuring the median or mode of the flux in small pixel regions over the entire detector, except inside the illuminated traces, and then fitting a smooth 2D function such as a 2D polynomial or a 2D spline. The best-fit model is then subtracted from the raw frame to remove stray light contamination. We note that all frames used to build the master flat-field calibration described above should be corrected for stray light.

\newpage

\section{Radial Velocity Modeling Packages}

A number of publicly available exoplanet fitting software packages exist, many of which support fitting not only the planet’s RV parameters but also some form(s)
of stellar variability, related transit photometry data, and even the stellar parameters of the host star. The table below contains an inexhaustive list of such packages with pointers to their reference articles and brief notes whether/how each package addresses fitting stellar variability and joint RV + transit photometry fits.

\begin{table}[h]
\tabcolsep7.5pt
\caption{A subset of publicly available radial velocity fitting packages}
\label{tab:rvfit_list}
\begin{center}
\begin{tabular}{@{}l|c|c|c|c@{}}
\hline
Package Name & Primary  & Variability & Joint RV + & Reference \\
     & Language & Fitting     & Transit?\\
\hline
EXOFASTv2  & IDL         & N/A      & Yes  & \citealt{Eastman2019} \\
Exoplanet  & Python      & GP       & Yes & \citealt{Foreman-Mackey2021} \\
Juliet     & Python      & GP       & Yes & \citealt{Espinoza2019} \\
Kima       & C++/Python  & Multi-Dim GP       & No  & \citealt{Faria2018} \\ 
Pexo       & R           & No        & No  & \citealt{Feng2019} \\ 
Pyaneti	   & Python/Fortran & Multi-Dim GP  & Yes  & \citealt{Barragan2022a} \\
PyOrbit	   & Python & Multi-Dim GP  & Yes & \citealt{Malavolta2016} \\
RadVel     & Python & GP            & No  & \citealt{Fulton2018} \\
\hline
\end{tabular}
\end{center}
\end{table}

\bibliographystyle{ar-style2}
\bibliography{references}

\end{document}